\documentclass[12pt]{article}
\pdfoutput=1
\usepackage{graphicx}
\title{BOOK: \\{\sc Models of science dynamics - encounters between complexity theory and 
information sciences} \vskip2cm
  \vskip2.5cm CHAPTER 3 \\    \vskip2.5cm
Knowledge epidemics \\and population dynamics models\\ for describing  idea
diffusion  } \vskip2.5cm
\author{\sc Nikolay K. Vitanov and Marcel R. Ausloos}
\date{}
\begin{document}
\pagestyle{headings}
\maketitle
\newpage
\pagestyle{plain}
\pagenumbering{arabic}
\begin{abstract}
The diffusion of ideas is often  closely connected to the creation and diffusion of 
knowledge and to the technological evolution of society.  Because of this,
knowledge  creation, exchange and  its subsequent  transformation into 
innovations for improved welfare and economic growth  is 
briefly described from a 
historical point of view. Next,   three approaches are discussed for modeling
  the diffusion of ideas in the areas of science and technology, through  
(i) deterministic, (ii) stochastic, and
(iii) statistical approaches. These are illustrated  through their corresponding 
population dynamics and epidemic models  relative to the spreading of 
ideas, knowledge and innovations. 
\par
The   deterministic dynamical models are considered to be
appropriate for analyzing the evolution of  large and small societal, scientific 
and technological systems
when the influence of fluctuations is insignificant.  Stochastic models are 
appropriate when the system of interest  is small but when the fluctuations become 
significant for its 
evolution. Finally   statistical approaches and models  based on the laws 
and distributions of Lotka, Bradford,  Yule,
Zipf-Mandelbrot, and others, provide much useful information for the analysis 
of the evolution of systems  in which development is closely connected to  the 
process of idea diffusion.
\end{abstract}
\newpage
\begin{table}[t]
\vskip-3.5cm
\begin{center}
\begin{tabular}{|c|c|}
\hline
{\sc 10 important  questions }  & {\sc and their answers} \\
{ \sc raised in  this chapter} & in the form of guidance \\
\hline\hline
1. What is the connection between  &Knowledge is often considered \\
knowledge and capital? & as a form of human capital  \\
\hline
2.  What happens in the case of   & Knowledge is  transferred \\
 knowledge diffusion? & when the subjects interact \\
\hline
3.  Should quantitative research be & Yes, surely supplemented \\
supplemented by qualitative research? &  coordinated joint aims   are useful  \\
\hline
4. Who are the pioneers of  & Alfred Lotka and \\
 scientometrics?  & Derek Price \\
\hline
5. What is the relation between epidemic & Epidemic models are  a\\
models and of & particular case of  \\
population dynamics models?   & population dynamics models \\
\hline
6. What has to be done if   & Switch from deterministic \\
fluctuations  strongly  influence &   to stochastic models \\
the system evolution?&     and think          \\
\hline
 & Often data is collected\\
 7. Why are discrete models useful?  & for some period of time. Thus, such data is \\
  & best described  by discrete models \\
\hline
8. Around which statistical law are  & \\
  grouped all statistical tools & Around   Lotka  law \\ 
  described in the chapter? & \\
\hline
9. Are all possibly relevant  models, & NO ! Only an appropriate selection. \\
   & For more models, consult the  literature  \\ 
 presented     in this chapter?       &   or ask a specialist \\ 
\hline 
& Proceed from simple to more \\
10.  What is the strategy followed& complicated models and from deterministic \\
 by the authors of the chapter?   & to stochastic models supplemented \\
  &  by statistical tools \\
\hline
\end{tabular}
\end{center}
\caption{Several questions and answers that  should guide and supply useful and important information
for the reader.}
\end{table}
\newpage
\begin{table}[t]
\vskip-3.5cm
\begin{center}
\begin{tabular}{|c|c|}
\hline
&\\
{\sc Models described in this  chapter}  & {\sc   are useful for} \\
&\\
\hline          \hline
 & Evaluation of research strategies. \\
     Science landscapes              & Decisions about personal development \\
                   & and promotion \\
\hline
Verhulst    & Description of  a large class of \\
    Logistic curve                &  growth processes \\
\hline
Broadcasting model      & Understanding the influence of mass \\
of technology diffusion & media on technology diffusion \\
\hline 
      &   Understanding the  influence of  \\
Word-of-mouth model              &   interpersonal contacts on \\
                   &   technology diffusion \\
\hline
   &    Understanding the  influence of both mass \\
Mixed information source model      &    media and interpersonal contacts on\\
                  &     technology diffusion \\
\hline
Lotka-Volterra model  &    Understanding the influence of the time lag \\ 
of innovation diffusion& between hearing about innovation and \\
 with time lag & its adoption \\
\hline
Price model of knowledge & Modeling the growth of discoveries, \\
growth with time lag     & inventions, and scientific laws \\
\hline
SIR models of scientific & Modeling the epidemic stage of \\
epidemics                & scientific idea spreading  \\
\hline
  & Extends the SIR model  \\
SEIR models of scientific                &  by specifically adding  the role \\
                   epidemics       & of a class of scientists exposed to \\
                    & some scientific idea\\
\hline
\end{tabular}
\end{center}
\caption{List of  models described in the chapter with  comments on
their usefulness.}
\end{table}
 
\newpage
\begin{table}[t]
\vskip-3.5cm
\begin{center}
\begin{tabular}{|c|c|}
\hline
&\\
{\sc Models described in this  chapter}  & {\sc   are useful for} \\
&\\
\hline  \hline
Discrete model for      &   Modeling and forecasting  \\
the change in the number of     &   the evolution in the number of \\
authors in a scientific  field&   authors and papers in a scientific field \\
\hline
Daley model             &   Modeling the evolution of a population \\    
                        &   of papers in a scientific field \\
\hline
Coupled discrete model  &   Modeling and forecasting the joint  \\
for populations of      &    evolution of population of scientists  \\
scientists and papers   &   and papers in a research field\\
\hline
   &   Epidemic model for  the increase of \\
Goffman-Newill model  &   number of scientists from a \\
for  the joint evolution of &   research field who start work \\
one scientific field and one      &   in a sub-field of the scientific  field.  \\
 of its sub-fields                         &  The model also describes   \\
                        &  the  increase in the number of papers in \\
                        &   the research sub-field \\ 
\hline
Bruckner-Ebeling-Scharnhorst & Understanding the  joint evolution\\
model for the  evolution of $n$   & of scientific fields in presence\\
scientific fields            & of migration of scientists from\\
                             & one field to another field\\
\hline
\end{tabular}
\end{center}
\caption{List of  models described in the chapter  with   comments on 
their usefulness (Continuing Table 2).}
\end{table}

\newpage
\begin{table}[t]
\vskip-3.5cm
\begin{center}
\begin{tabular}{|c|c|}
\hline
&\\
{\sc Models described in this  chapter}  & {\sc   are useful for} \\
&\\
\hline   \hline
SI model for the probability of& Modeling the spread of intellectual\\
intellectual infection     & infection along a scientific network\\
\hline 
& Modeling  the spread of intellectual infection\\
SEI model for the probability of   & along a scientific network in \\
  intellectual infection                            & the presence of a class of scientists \\
                            & exposed to the intellectual infection\\
\hline
 & Modeling the number of scientists in a \\
 Stochastic evolution model                          & research subfield as a stochastic   \\
                           & variable described by a master equation\\
\hline 
       & Modeling the influence of fluctuations    \\
 Stochastic model of  &  in scientific productivity    \\
 scientific productivity                            & through differential equations for \\
                           & the dynamics of a scientific community\\ 
\hline
Model of competition & Understanding the  competition between\\
 between ideologies                  & ideologies with possible \\
                            & migration of believers \\
\hline
Reproduction-transport      & Modeling  the change of research field\\
model                       & as a migration  process\\ 
\hline
\end{tabular}
\end{center}
\caption{List of  models described in the chapter with   comments on 
their usefulness (Continuation of Table 2).}
\end{table}
 
\newpage
\begin{table}[t]
\vskip-3.5cm
\begin{center}
\begin{tabular}{|c|c|}
\hline
&\\
{\sc Laws described in this  chapter}  & {\sc   are useful for} \\
&\\
\hline  \hline
& Describing the number  distribution  \\
 Lotka  law                                          &of  scientists with respect to \\
                           &the  number of  papers they wrote\\
\hline
   &  Writing \\ 
 Pareto    distribution  &   a  continuous version  \\
                           & of Lotka  law \\
\hline
Zipf law    &      Ranking scientists  \\ 
       and  &  by the number of papers
       \\ Zipf-Mandelbrot law   
                    &      they wrote\\
\hline
    &      Reflecting the fact that a large number\\
  Bradford law                    &      of relevant articles are concentrated\\
                    &      in  a small number of journals\\ 
\hline
\end{tabular}
\end{center}
\caption{List of  laws discussed in the chapter with  a few  words on 
their usefulness (Continuation of Table 2).}
\end{table}
\section{Knowledge, capital, science research, and  ideas diffusion}
\subsection{Knowledge and capital}
Knowledge can be defined as a dynamic framework connected to cognitive structures 
from which information can be sorted, processed and understood \cite{howells02}.  
Along  economics lines of thought \cite{barro,leyd,dolf}, knowledge  can be treated 
as one of the "production factors", - i.e., one of the main causes 
of wealth in modern 
capitalistic societies.  

According to 
Marshall  \cite{marsh} a "{\bf  capital}" is a collection of goods
external to the economic agent that can be sold for money and from which  an income
can be derived.  Often, knowledge is  parametrized as such a  "{\bf  human capital}" 
\cite{romer,romer1,romer2,romer3,jaf1}.
Walsh \cite{walsh}  was one pioneer  in  treating human knowledge
as if it was  a "capital", in the economic sense; he  made an attempt to find   measures for this form
of  "capital".  Bourdieu \cite{bourdieu}, Coleman \cite{coleman}, Putnam 
\cite{putnam}, Becker and collaborators have further  implanted the concept of   
such a "human capital" in   economic theory \cite{beck1,beck2,stiglitz}. 
\par
However,  the concept of knowledge as a form of capital is an oversimplification. 
This   global-like concept does not  account for many properties of knowledge  
strictly connected to the individual,  such as the possibility for different 
learning paths or different views,  multiple levels of 
interpretation, and different preferences \cite{davis}. In fact, knowledge develops in a  quite
complex social context,  within possibly different frameworks or time scales, and involves "tacit  dimensions"  (beside the basic space and time  dimensions) requiring coding and 
decoding \cite{dolf}. 
\begin{center}
{\fbox{\fbox{\parbox{12.5cm}{ \vskip0.5cm \sf 
FOR POLICY-MAKERS\\
 Take away box Nr.1: Knowledge is much more than a form of capital:it is a dynamic framework connected to cognitive structures 
from which information can be sorted, processed and understood.
 \vskip0.5cm }}}}\\
\end{center}
\subsection{Growth and exchange of knowledge}
Science policy-makers and scholars have for many decades wished  to develop  quantitative 
methods for describing  and predicting  the initiation and growth of science research
\cite{price4,price3,foray}.  Thus, scientometrics  has become  
one of the core research activities    in  view  of constructing science and technology indicators
\cite{raan}.
\par    
The accumulation of the knowledge in  a country's population arises either from acquiring 
knowledge from abroad  or from  internal engines \cite{non1, non2, non3, bernx}.  The main  
engines for the production 
of new knowledge in  a country are usually:  the public research institutes,   the universities 
and training institutes, the firms, and  the individuals  \cite{dahlman}. The users 
of the knowledge are  firms, governments, public institutions (such as the national education, 
health, or security institutions), social organizations, and any concerned individual. The knowledge is 
transferred from producers to the users by dissemination  that is realized by some
flow or diffusion of  process \cite{dahl2}, sometimes involving physical migration. 
\par
Knowledge typically appears at first as purely tacit: {\it a person "has" an idea} \cite{saviotti99,cowan97}. 
 This  tacit knowledge must be codified for further use;  after codification, knowledge 
can be stored in  different ways,  as in textbooks or digital carriers. It
 can be transferred from one system to another. In addition to knowledge creation, a system can gain
knowledge by knowledge exchange and/or   trade. 
\par
In   knowledge diffusion, the  knowledge is
transferred while subjects interact  \cite{jaff86,antonelli96,morone2010}. Pioneering studies on knowledge  diffusion investigated the patterns 
through which new technologies are spread in social systems \cite{rogers62,casseti69}. The gain of knowledge due to knowledge diffusion is one of the keys or leads to innovative products 
and innovations \cite{kuch, ax2}.
\begin{center}
{\fbox{\fbox{\parbox{13.0cm}{ \vskip0.5cm \sf 
FOR POLICY-MAKERS\\
 Take away box Nr.2: \\ An innovative  product or a process is {\bf new} for 
the group of people who are 
likely to use it. Innovation is an  innovative product or process that has
passed the barrier of user adoption. Because of the rejection by the market, 
many innovative products and processes never become an innovation. 
 \vskip0.5cm }}}}\\
\end{center} 
\par
In science,  the  diffusion  of knowledge is  mainly connected to the transfer of scientific 
information by publications. It is accepted that the results of some research become 
completely scientific when they are published \cite{ziman}. Such a diffusion can also take place at scientific meetings and  through oral or other exchanges, sometimes without formal publication of exchanged  ideas \cite{gordon}.
\begin{center}
{\fbox{\fbox{\parbox{13.0cm}{ \vskip0.5cm \sf 
FOR POLICY-MAKERS\\
 Take away box Nr.3: \\ Scientific communication
has specific features. For example,  citations are very important in the communication
process as they place corresponding research and researchers, mentioned in the scientific literature,  in a way 
similar to the kinship links that  tie persons within a tribe.  
Informal exchanges   happening in the process of common work at the
time of meetings, workshops, or conferences may accelerate the transfer of scientific
information,  whence the growth of knowledge   
 \vskip0.5cm }}}}\\
\end{center} 
\section{Qualitative research.  Historical remarks.}
\subsection{Science landscapes}
Understanding   the diffusion of knowledge requires research complementary
to  mathematical investigations. For example, mathematics cannot indicate why 
the exposure to  ideas leads to intellectual epidemics. Yet, mathematics can provide 
information  on the intensity or the duration  of  some 
intellectual epidemics.  

Qualitative research is all about exploring issues, 
understanding phenomena, and answering questions \cite{Brymanbook} without much mathematics. 
Qualitative research involves     empirical research  through  which the researcher explores 
relationships using a textual methodology 
rather than quantitative data. Problems and results in the field of qualitative 
research on knowledge epidemics will not be discussed in detail here.  
However,  through one example it can be shown   how mathematics can create the basis for qualitative research 
and decision making. This example is connected to the science 
landscape concepts outlined here below.
\par 
The idea of science landscapes has some similarity with the work  of Wright \cite{wright} in 
biology who proposed that 
the fitness landscape evolution can be treated as as optimization process  based on the roles of mutation, inbreeding, crossbreeding, and selection.
The science landscape idea was    developed by Small \cite{s1,s2}, 
 as well as by Noyons and van Raan  
\cite{noyons}. In this framework,  Scharnhorst \cite{a2,a1} proposed  an approach
for the analysis of scientific landscapes, named  "geometrically oriented evolution theory".
\begin{center}
{\fbox{\fbox{\parbox{13.0cm}{ \vskip0.5cm \sf 
FOR POLICY-MAKERS\\
 Take away box Nr.4: \\The concept of science landscape is rather simple:  Describe the 
corresponding field of
science or technology through a function of parameters such as height, weight, size, 
technical data, etc. Then a  virtual knowledge landscape 
can be constructed from empirical data in order  to visualize and understand 
innovation and  to optimize various processes in science and technology. 
 \vskip0.5cm }}}}\\
\end{center} 
As an illustration at this level, consider that a mathematical  example of a technological landscape  can be  given by a function
$C=C(S,v)$,  where $C$ is the cost for developing  a new airplane, and where $S$ and $v$ represent the 
size and velocity of the airplane. 
\par
Consider  two
examples concerning the use of science landscapes for evaluation purposes:

{\bf (1)  Science landscape  approach as a method for  evaluating national research strategies } 

 For  example,  
national science systems can be considered as made of researchers who 
compete for scientific results,  and subsidies,  following optimal research strategies. The efforts of every country become visible, comparable and measurable
by means of   appropriate functions or landscapes: e.g.,  the number of 
publications. The aggregate research strategies of a country  can thereby be represented 
by the distribution of publications in the various scientific 
disciplines. In so doing, within a two-dimensional space,\footnote{
E.g., take the scientific 
disciplines and the number of publications as axes} different countries correspond to different landscapes. 
Various political discussions  can follow and evolution strategies can be invented thereafter. 
\par
Notice that the dynamics of self-organized
structures in complex systems can be understood as the result of a search for 
optimal solutions to  a certain problem. Therefore, such a comment shows how rather strict mathematical approaches,  not disregarding simulation methods, can be congruent to qualitative questions.
\par
{\bf (2)  Scientific citations as landscapes  for  individual evaluation} \\
Scientific citations  can serve for constructing landscapes. Indeed, citations have a key position in the retrieval and valuation of 
information in scientific communication systems \cite{a2,egg,egg1}. 
This position is based on the
objective nature of the citations as components of a global information system, as represented 
by the Science Citation Index. A landscape function  based on citations can be defined in various ways. It can take into account self-citations \cite{iina1,iina2,iina3,pekalski}, or  time-dependent quantitative measures \cite{hindex,soler,bur07}.

\begin{center}
{\fbox{\fbox{\parbox{13.0cm}{ \vskip0.5cm \sf 
FOR POLICY-MAKERS\\
 Take away box Nr.5: \\ Citation landscapes become important elements  of 
a science policy (e.g.,  in personnel management decisions), thereby influencing 
individual scientific careers,  evaluation of research institutes, and 
investment strategies.
 \vskip0.5cm }}}}\\
\end{center}  
\subsection{Lotka and Price: pioneers of  scientometrics}
Alfred 
Lotka, one of the modern founders of  population dynamics studies, was also an excellent
statistician. He discovered \cite{lotka26} a distribution for the number of authors 
$n_r$  as a function of the  number of published papers $r$, - i.e., $n_r = n_1/r^2$. 
\par
However, Derek Price, a physicist, set the mathematical 
basis    in the field of measuring scientific research in recent times \cite{price63,price75,price78}.  
He proposed a model of scientific growth connecting science and 
time. In the first version of the model, the size of science was measured
by the number of  journals founded  in the course of  a number of years. Later,  instead of the number 
of journals, the number of published  papers
was used as the measure of scientific growth. Price and other authors 
\cite{price75,price78,gilbert} considered also different indicators of scientific growth, 
such as the number of authors, funds, dissertation  production, citations, or the number of scientific 
books.

In addition to the deterministic approach   initiated by 
Price,
the statistical approach to the study of scientific information developed rapidly and  nowadays  is still  an  
important tool in   scientometrics   \cite{chung,kealey}. More  discussion on the statistical approach will be given in section 6
of this chapter.

\begin{center}
{\fbox{\fbox{\parbox{13.0cm}{ \vskip0.5cm \sf 
FOR POLICY-MAKERS\\
 Take away box Nr.6: \\
 Price distinguished three stages in the growth of knowledge: 
{\bf (a)} a preliminary phase with small increments; {\bf (b)} a phase of exponential 
growth; {\bf (c)} a saturation stage. The stage (c) must be reached sooner or later  
after the new ideas and  opportunities are exhausted; the growth slows down
until a new trend emerges and gives rise to a new growth stage. According to
Price, the curve of this  growth is a S-shaped logistic curve. 
 \vskip0.5cm }}}}\\
\end{center} 
\subsection{Population dynamics and   epidemic models of the diffusion of knowledge}
\begin{figure}[h]
\begin{center}
\includegraphics[scale=0.5]{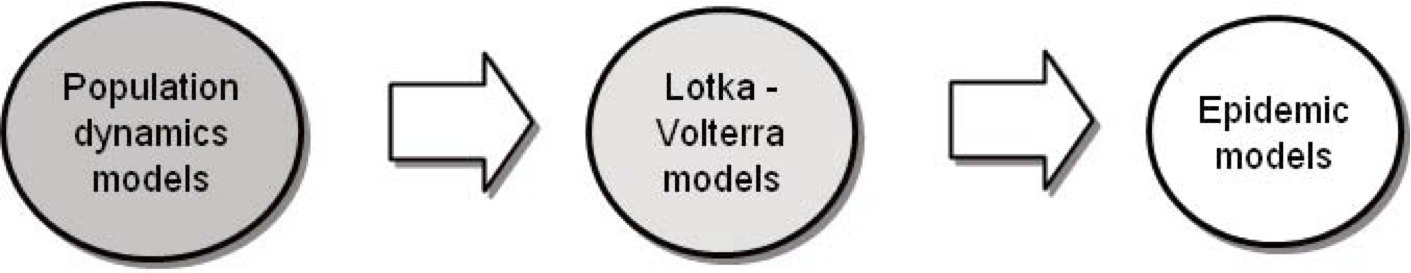}
\end{center}
\caption{Relation among epidemic models, Lotka-Volterra models, and 
population dynamics models.}
\end{figure}
\par
Population dynamics is the branch of life sciences that studies short- and long-term changes 
in the size and age composition of populations, and how the biological and environmental 
processes influence those changes. In the past, most  models  for 
 biological population dynamics have been of interest only  in   
mathematical biology \cite{murray,edelst}.  Today, these models are adapted and applied in 
many more areas of science \cite{dietz, dodd}. Here below, models of knowledge 
dynamics  will be of  interest as bases of epidemic models. Such models are nowadays used  
because   some stages of idea spreading   processes  within a population (e.g,  of 
scientists),   possess properties  like those of epidemics.
\par
The mathematical modeling of epidemic processes has attracted much attention since  the 
spread of infectious diseases has always been of great concern and considered to be  a threat 
to public health \cite {anderson,brauer,ma}. In the history of science  and 
society,  many examples of ideas spreading seem to occur in a way similar to the spread of epidemics.  Examples of the former 
field pertain  to the ideas  of Newton on mechanics and the passion for "High Critical 
Temperature Superconductivity" at the end of the twentieh century. Examples  of the latter field 
are the spreading of ideas from Moses or Buddha \cite{gof2},  
or  discussions based on the Kermack-McKendrick model \cite{kmk}  for the epidemic stages 
of  revolutions  or  drug spreading \cite{epstein}.
\par
Epidemic models belong to  a more general class of Lotka-Volterra models  used in   
research  on systems  in the fields of biological population dynamics,  
social dynamics, and economics. The models can also be used for describing  processes connected to the 
spread of knowledge, ideas  and innovations (see Fig. 1). Two examples are the  model of 
innovation in established organizations \cite{cast}
and  the Lotka-Volterra model for forecasting emerging technologies and  the growth of
knowledge \cite{kuch}. In   social dynamics, 
the Lanchester model of war between two armies can be mentioned, a model which in the case of 
reinforcements coincides with the Lotka-Volterra-Gause model for competition between two 
species \cite{gause}. Solomon and Richmond \cite{sr,sr1}  
applied  a Lotka-Volterra model  to financial markets,     while the model for the trap of 
extinction  can be applied to  economic subjects \cite{dv06}. Applications to
chaotic pairwise competition among political parties \cite{dv04} could also be mentioned.
\begin{figure}[h]
\begin{center}
\includegraphics[scale=0.5]{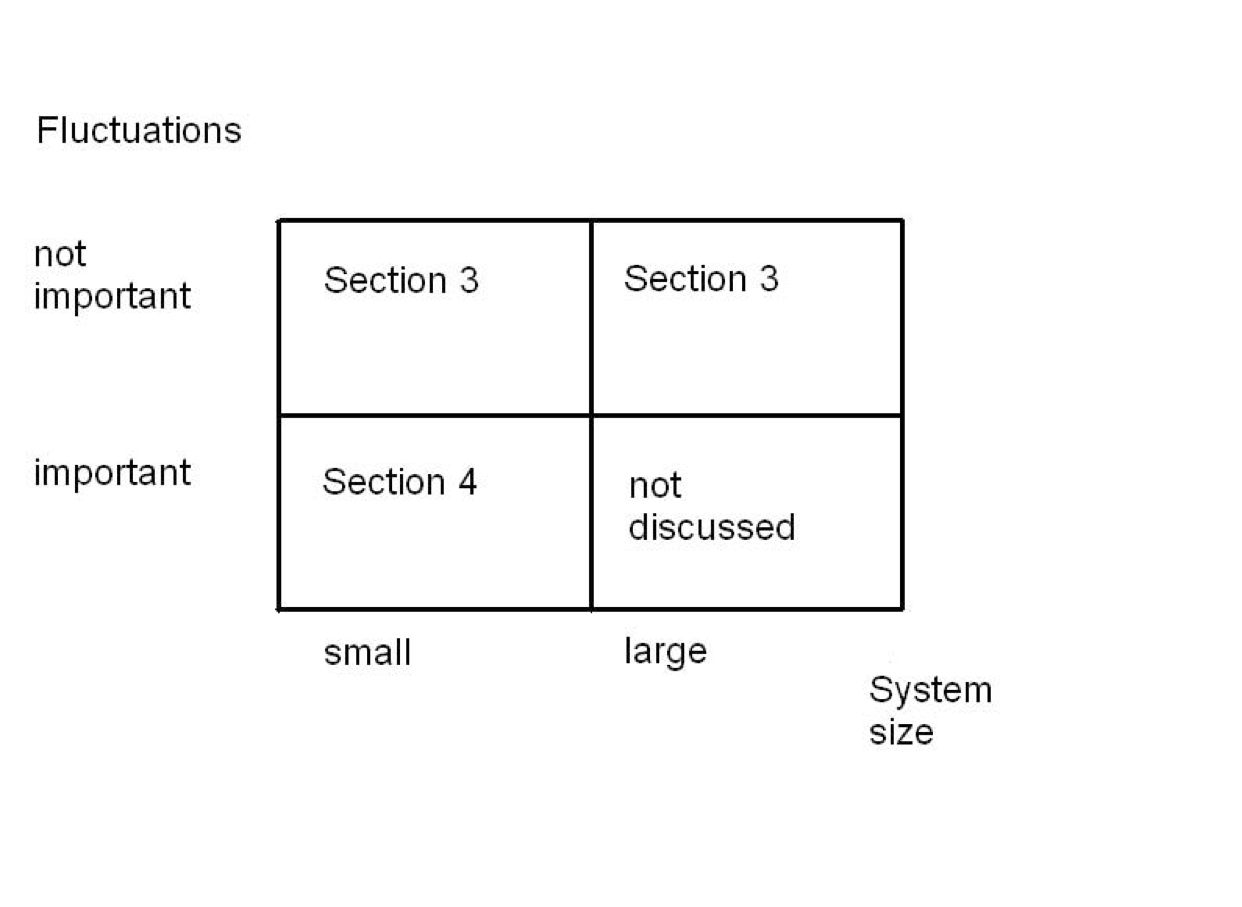}
\end{center}
\caption{Relationships between system size, influence of fluctuations,
and discussed classes of models.}
\end{figure}
\par
To start the discussion of population dynamics models  as applied to the growth of
scientific knowledge with special emphasis on  epidemic models,
two kinds of models can be discussed (Fig. 2): {\bf(1) deterministic models},  see Sec. 3, appropriate for 
large and small populations where the fluctuations are not drastically important,  
{\bf (2) stochastic models}, see Sec. 4,  appropriate for small populations. In the latter case  the 
intrinsic randomness  appears much more relevant than in the former  case.  Stochastic models for large populations will not be discussed. The reason for this is that such models 
usually consist of many stochastic differential equations, whence  their evolution can be 
investigated only numerically.
\par
Finally, let us mention that the knowledge diffusion is closely connected to the structure
and properties of the social network where the diffusion happens. This is a new and very promising
research area. For  example, a combination can be made between the theory of information
diffusion and the theory of complex networks \cite{bocc}. For more information about the
relation between networks and knowledge, see the following chapters of the book.
\section{Deterministic models}  
\begin{figure}[h]
\begin{center}
\includegraphics[scale=0.4]{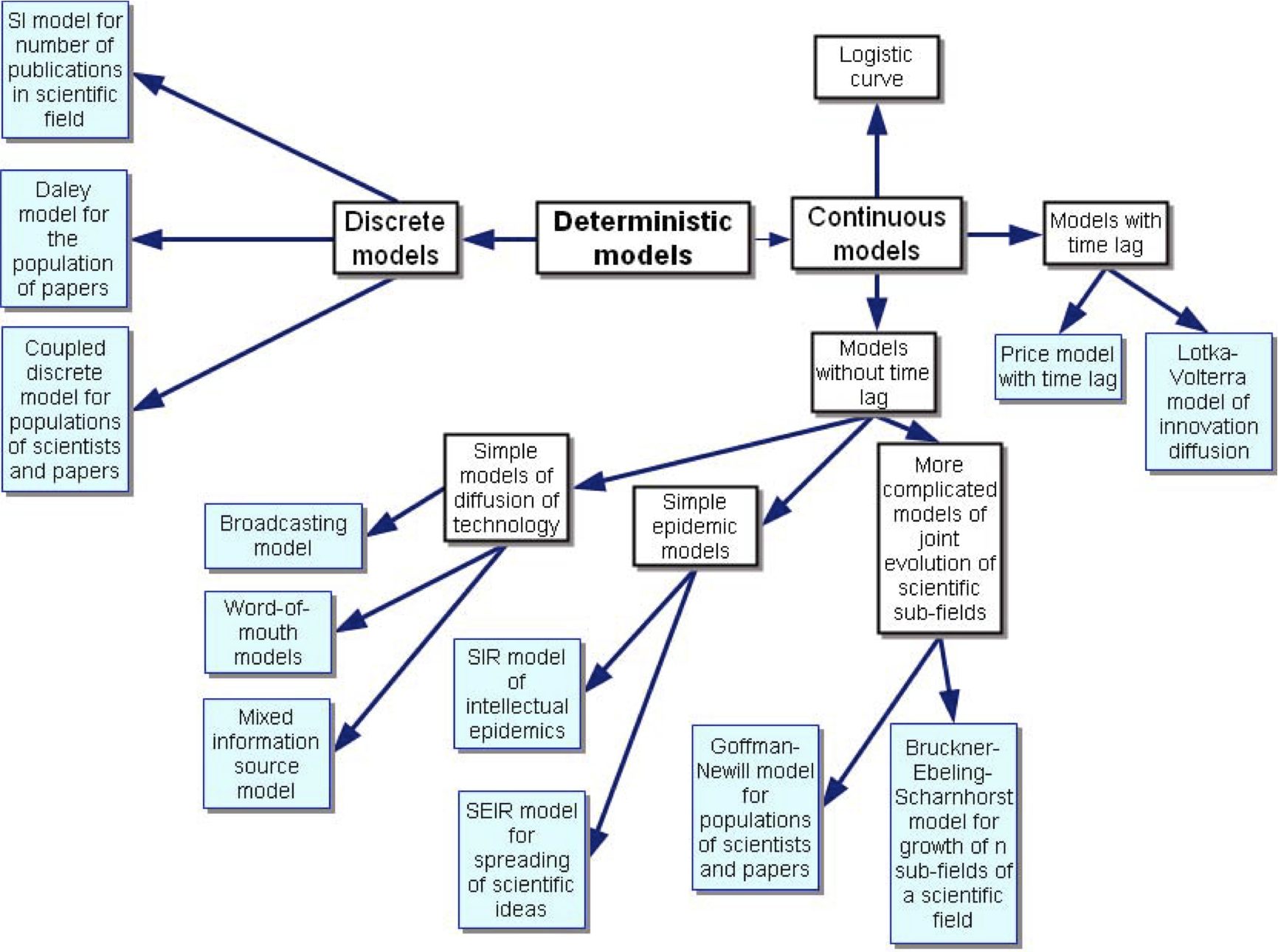}
\end{center}
\caption{Discrete (3) and  continuous (10) models discussed in the chapter.
Two continuous models account for
the influence of time lag, three models are simple models of 
technological diffusion. Two models are
simple epidemic models and two  models are  more complicated models. 
In addition, the basic logistic curve is discussed.}
\end{figure}
Below, 13 selected deterministic models (see Fig. 3) are discussed. The emphasis is 
on  models that can be used  for describing the epidemic stage of the diffusion of ideas, 
knowledge, and technologies. 
\subsection{Logistic curve and its generalizations}
In a number of cases, the natural growth of autonomous  systems in competition can be 
described by the logistic equation and the logistic curve  (S-curve)  \cite{meyer}.   In 
order to describe trajectories of growth or decline
in socio-technical systems, one generally applies a three-parameter logistic curve:
\begin{equation} \label{Scurve}
 N(t) = \frac{K}{1+ \exp[-\alpha t - \beta]}  
\end{equation}
where $N(t)$ is  the number of units in the 
species or growing variable to study; $K$  is the asymptotic limit of growth; $\alpha$  is 
the growth rate which specifies  the "width"  of the S-curve for $N(t)$; and  $\beta$  specifies 
the time $t_m$ when the curve reaches the midpoint of the growth trajectory,  such that  $ 
N(t_m) = 0.5 \;K$. The three parameters, $K$, $\alpha$, and $\beta$, are usually  obtained after 
fitting  some data \cite{meade}. It is well known that many cases of epidemic growth can be 
described by parts of an appropriate S-curve.  As an example, recall that the S-curve was 
also used for 
describing  technological substitution  \cite{rogers62,mans,modis}, $ca.$ 60 years ago.
\par
However, different interaction schemes  can generate different  growth patterns for  whatever  
system species are under consideration \cite{modis2}. Not every interaction scheme
leads to a logistic growth \cite{maCHESS}.  The evolution of systems in such regimes may be 
described by more complex curves, such as a combination of 
two or more simple three-parameter functions  \cite{meyer,meyer2}. 
\subsection{Simple epidemic and Lotka-Volterra models of technology diffusion}
As recalled here above, the simplest epidemic models could be used for describing technology 
diffusion, like  considering  two populations/species: adopters and non-adopters of some 
technology. Such models can be put into   two basic classes:  either broadcasting (Fig. 4) or 
word-of-mouth models (Fig. 5). In the 
broadcasting models, the source of knowledge about the existence and/or characteristics of
the new technology is external and reaches all possible adopters in 
the same way. In the word-of-mouth models, the knowledge is diffused by means of 
personal interactions.
\par
{\bf (1) The broadcasting model (Fig. 4)}\\
\begin{figure}[h]
\begin{center}
\includegraphics[scale=0.6]{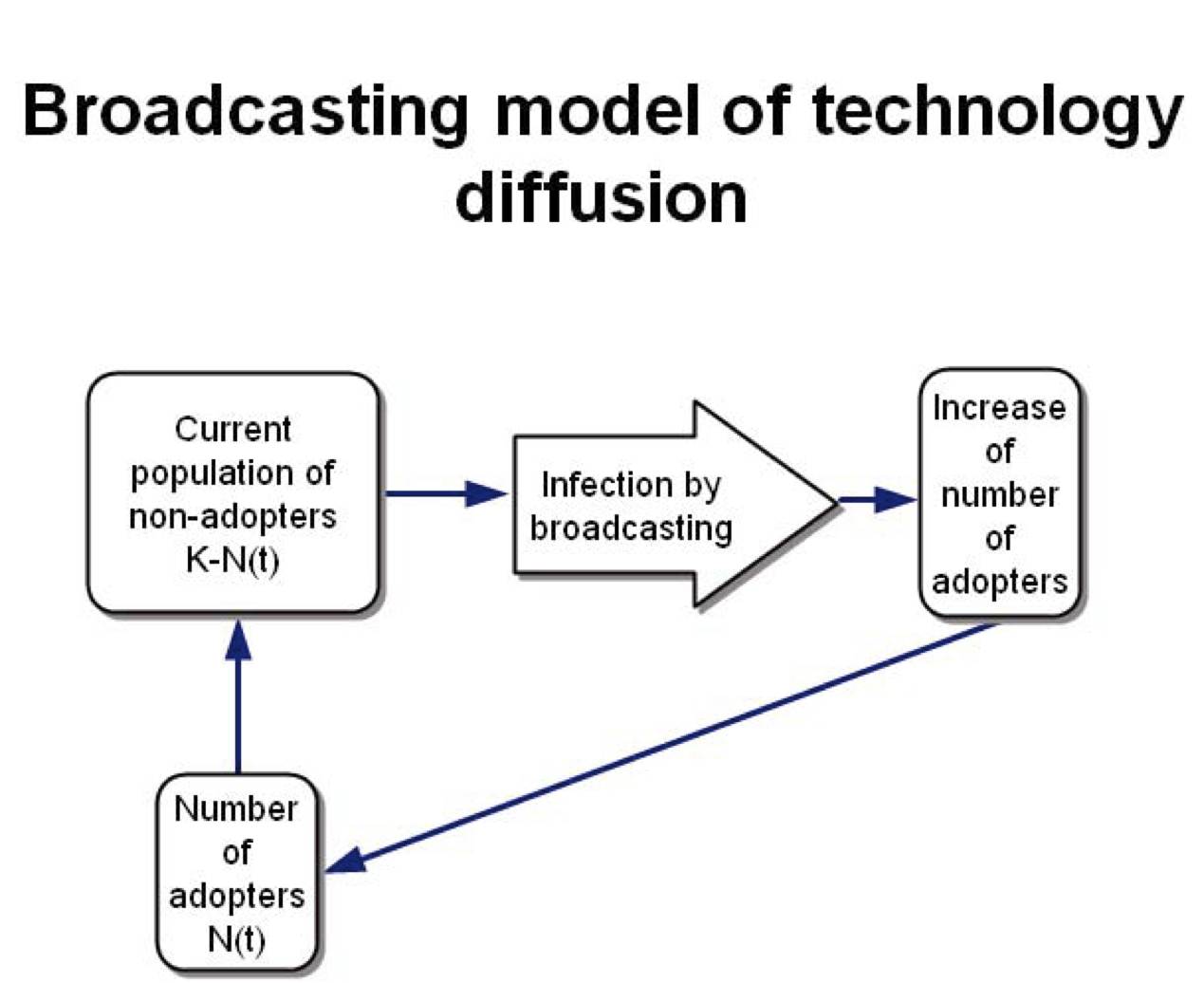}
\end{center}
\caption{Schematic representation of a broadcasting model of technology
diffusion. The number of adopters of technology increases by mass
media influence.}
\end{figure}
Let us consider a population of $K$ potential adopters of the new technology and
let each adopter  switch to the new technology as soon as he/she hears 
about its existence (immediate infection through  broadcasting). The probability that at time $t$ 
a new subject 
will adopt the new technology is characterized by a coefficient of diffusion $\kappa(t)$ which
might or might not be a function of the number of previous adopters.
In the broadcasting model $\kappa(t) = a$ with $(0 < a < 1)$;  this  is considered to be a 
measure of the  infection probability.
\par
Let  $N(t)$ be the  number of adopters at time $t$.
The increase in adopters for each period is equal to the probability
of being infected, multiplied by the current population of
non-adopters \cite{mah85}. The rate of diffusion at time $t$ is
\begin{equation}\label{eqg1}
\frac{d N}{d t} = a [K - N(t)].
\end{equation}
The integration of (\ref{eqg1}) leads to the number of adopters: i.e.,  \begin{equation}\label
{eqg1i} N(t) = K [1- \exp(-at)]. \end{equation} $N(t)$  is described by a decaying 
exponential  curve.
\par
{\bf (2) Word-of-mouth model (Fig. 5)}\\
\begin{figure}[h]
\begin{center}
\includegraphics[scale=0.5]{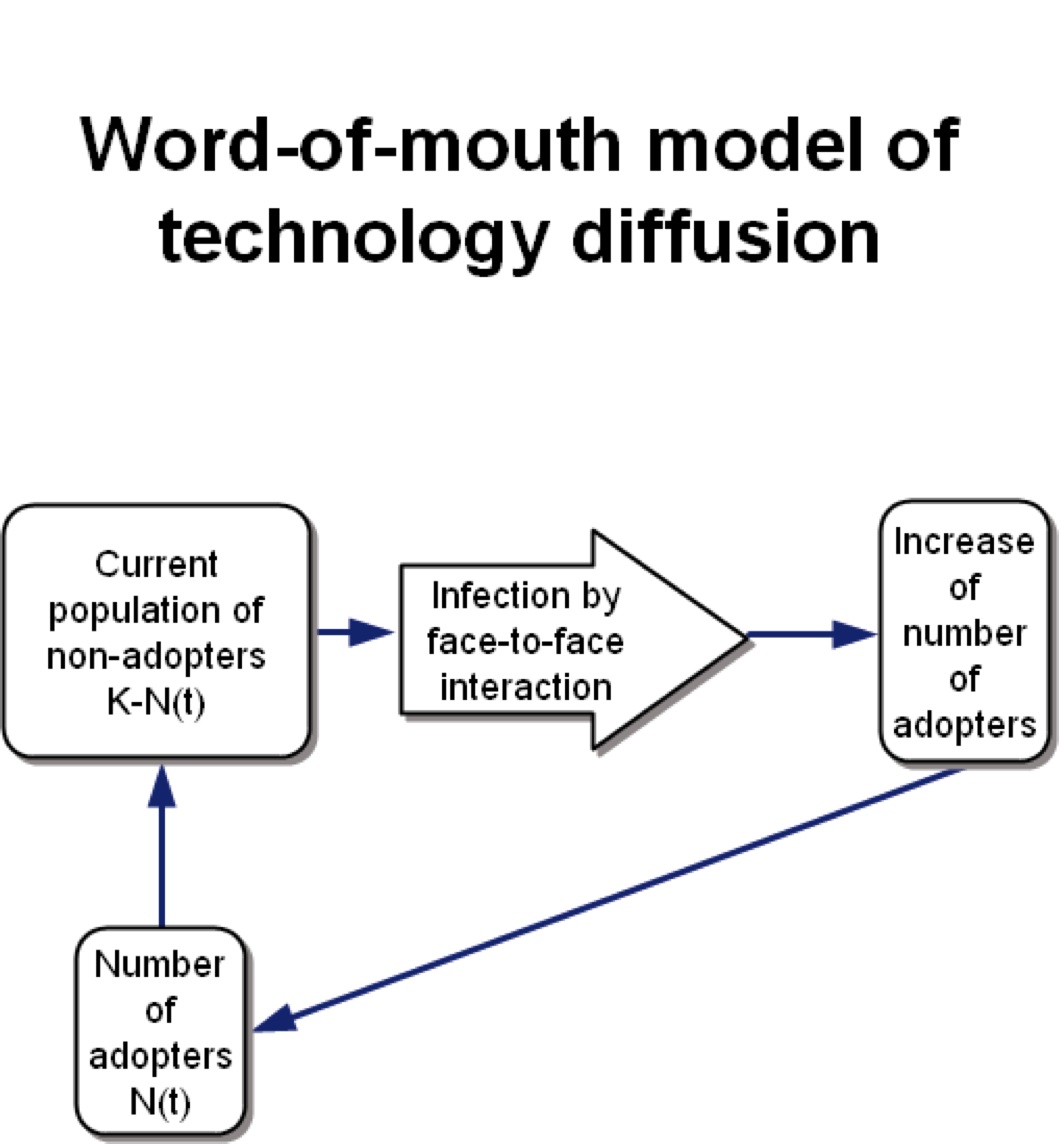}
\end{center}
\caption{Schematic representation of a word-of-mouth model of technology
diffusion. The number of adopters of technology increases by interpersonal
interactions.}
\end{figure}
In many cases, however, the technology adoption timing is at least an order of magnitude  
slower than  the time it takes for information  spreading  \cite{ger00}.
This  requires another modelization than in {\bf (1)}:  the word-of-mouth 
diffusion model. Its basic assumption is that knowledge diffuses by
means of face-to-face interactions. Then the probability of
receiving the relevant knowledge needed to adopt the new technology
is a positive function of current users $N(t)$. Let the 
coefficient of diffusion $\kappa(t)$ be $b N(t)$ with $b > 0$. The rate of diffusion 
at time $t$ is 
\begin{equation}\label{eqg3}
\frac{dN}{dt} = b\;N(t)\;[K - N(t)]\; .
\end{equation}
Then \begin{equation}\label{eqg3i}
N(t) = \frac{K}{1 + \left(\frac{K-N_0}{N_0} \right)e^{-b K(t-t_0)}} 
\end{equation}
where $ N_0=N(t = t_0) $. $N(t)$  is described by an S-shaped curve. 
\par 
A  constraint exists in the word-of-mouth  model:  it 
explains the diffusion of an innovation not from the date  of its invention but
from the date when some number, $N(t) > 0$, of early users have
begun using it.
\par
{\bf (3) Mixed information source model (Fig. 6)} \\ 
\begin{figure}[h]
\begin{center}
\includegraphics[scale=0.6]{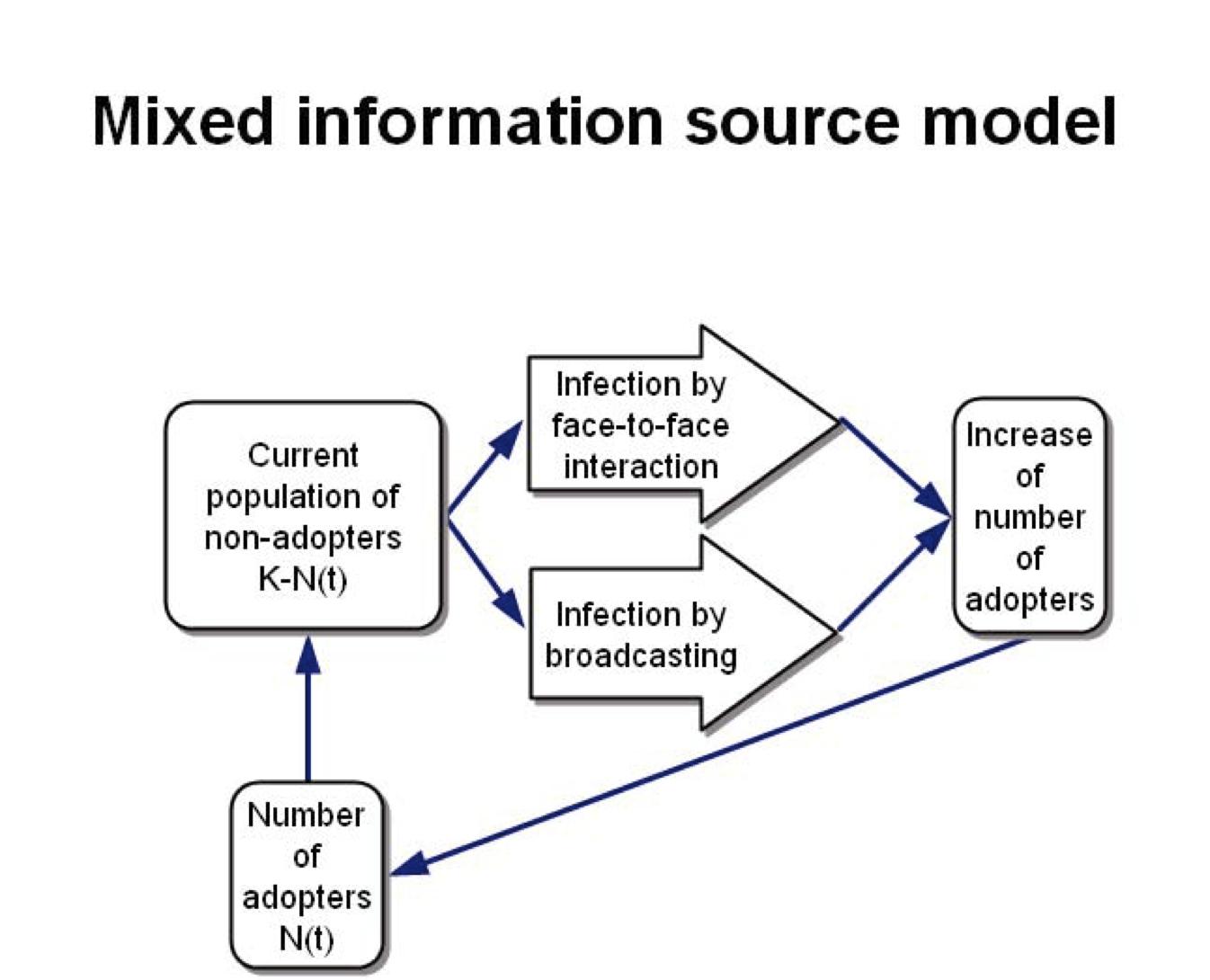}
\end{center}
\caption{Schematic representation of mixed information source model.
The number of adopters increases by mass media influence and interpersonal
contacts.}
\end{figure}
 In the mixed information source model, existing
non-adopters are subject to two sources of information (Fig. 6). The
coefficient of diffusion is  supposed to look like $a + b N(t)$. The  model evolution 
equation  becomes 
\begin{equation}\label{eqg5}
\frac{dN}{dt} = (a+b N(t))\;[K - N(t)].
\end{equation}
The result of Eq.(\ref{eqg5}) is a (generalized) logistic
curve whose shape is determined by  $a$ and $b$ \cite{mah85}.
\par
{\bf (4)  Time lag Lotka-Volterra model of innovation diffusion (Fig. 7)}\\
\begin{figure}[h]
\begin{center}
\includegraphics[scale=0.4]{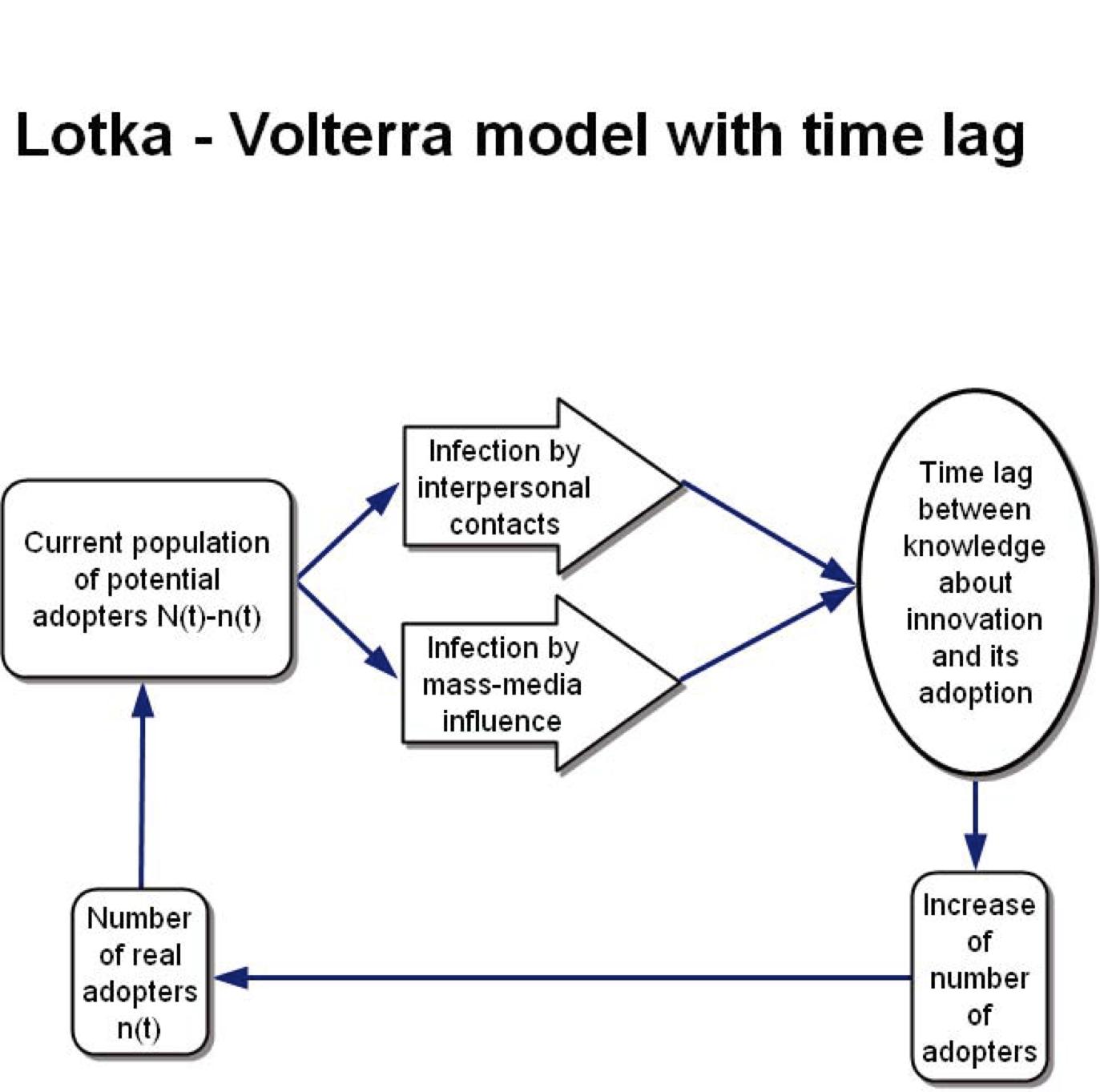}
\end{center}
\caption{Schematic representation of a Lotka-Volterra model with time lag.
The model accounts for the time lag between hearing about innovation
and its adoption.}
\end{figure}
Let it be again assumed that the diffusion of innovation in a society is accounted for by a 
combination of two processes: a mass-mediated process and a process connected to 
interpersonal  (word-of-mouth) contacts. Let $N(t)$  be the number of potential adopters. 
Some of the 
potential adopters adopt the innovation and become real adopters.  The equation for the the 
rate of growth of the real adopters $n(t)$, in absence of time lag, is
\begin{equation}\label{eqex1}
\frac{d n(t)}{dt} = \alpha [N(t) - n(t)] + \beta n(t) [N(t) - n(t)] - \mu n(t) , 
\end{equation}
where $\alpha$ denotes the degree of external influence such as mass media, $\beta$ accounts 
for the degree of internal influence by interpersonal contact between adopters and 
the remaining population;  $\mu$ is a parameter characterizing the decline 
in the number of adopters because of  technology rejection for whatever reason. 
\par
A basic limitation in most models of  innovation diffusion 
has been the assumption of instantaneous acceptance of the new innovation by a 
potential adopter \cite{mah85,bart}. Often, in reality, there is a finite time 
lag between the moment when a potential adopter hears about a new innovation and the 
time of adoption. Such time lags usually are continuously distributed \cite{may,lal}. 
\par
The  time lag between the knowledge  about the innovation and its adoption can be 
captured by a distributed time lag approach in which the 
effects of time delays are expressed as a weighted response over a finite time  interval 
through appropriately chosen memory kernels \cite{k82} (see Fig. 7). 
Whence Eq.(\ref{eqex1}) becomes
\begin{eqnarray}\label{eqex2}
\frac{dn(t)}{dt} = \alpha \int_0^t d \tau \ K^*_1(t-\tau) \ [N(\tau) - n(\tau)]
+ \nonumber \\
 \beta \int_0^t d \tau \ K^*_2 (t-\tau) n(\tau) [N(\tau) - n(\tau)] - 
\mu \int_0^t d \tau \ K^*_3(t- \tau) n(\tau).
\end{eqnarray}
\par
Eq.(\ref{eqex2}) reduces to Eq.(\ref{eqex1}) when the 
memory kernels $K^*_i(t)$ ($i= 1,2,3$) are replaced by delta functions.
\par
 Two generic types of 
kernels are usually considered \cite{lal}: 
\begin{eqnarray}\label{eqex22222}
 K^*(t) = \nu \; e^{- \nu t}  \\
 K^*(t) = \nu^2 t \;  e^{- \nu t} \; ,
\end{eqnarray}
in which $\nu^{-1}$  is  some characteristic  time scale of the system.  
\par
The number of potential adopters $N(t)$  changes over time. 
Several possible functional forms of $N(t)$ are used \cite{shar}:
\begin{equation}\label{eqex5}
N(t) = N_0 (1 + at); \hskip.5cm N_0 >0, a>0 \\
\end{equation}
\begin{equation}  \label{eqex6}N(t) = N_0 \exp[gt]; \hskip.25cm   N_0 >0, g>0
\end{equation}
\begin{equation}\label{eqex7}
N(t) = \frac{b}{1+ d\exp(-ct)}; \hskip.5cm b>0, d>0, c>0
\end{equation}
\begin{equation}\label{eqex8}
N(t) = b - q \exp(-rt); \hskip.5cm b>0, q>0, r>0.
\end{equation}
Eq.(\ref{eqex6}) represents an approximation for short- and medium-term forecasting 
since  for $t$ large, $N(t)$ grows without bound, as in Keynes \cite{Keynes}. Eqs.(\ref
{eqex7}) and (\ref{eqex8}) are useful in 
long-term forecasting as $N(t)$ has an upper limit. Such forms for $N(t)$ are valid 
within a deterministic framework. 
\par
However, a stochastic framework (see below)  is more appropriate 
when the carrying capacity $N(t)$ is governed by some stochastic process, as when the influence 
of  socioeconomic and natural factors   are subject to "random" or hardly explainable 
fluctuations. In such systems,  $N(t)$ can be time-dependent: for  
example, $N(t) \sim N_0 (1 + \epsilon \cos (\omega t))$ where $\epsilon <<1$ and the
periodicity takes into account the influence of some (strong) cyclic economic factors.
In presence of a strong stochastic component, $N(t)$ can be stochastic: $N(t) = N_0 + \xi (t)$, 
where the noisy component is $\xi (t)$ and $N_0$ is the average value of the so-called 
carrying capacity \cite{Odum59}.
\begin{center}
{\fbox{\fbox{\parbox{13.0cm}{ \vskip0.5cm \sf 
FOR POLICY-MAKERS\\
 Take away box Nr.7: \\Time lags  between observations and decisions lead to   complicated dynamics. 
 Perform some preliminary
careful analysis of system behavior based on time lags  before making a decision.  
 \vskip0.5cm }}}}\\
\end{center} 
\subsection{ Price model of knowledge growth. Cycles of growth of knowledge}
\begin{figure}[h]
\begin{center}
\includegraphics[scale=0.37]{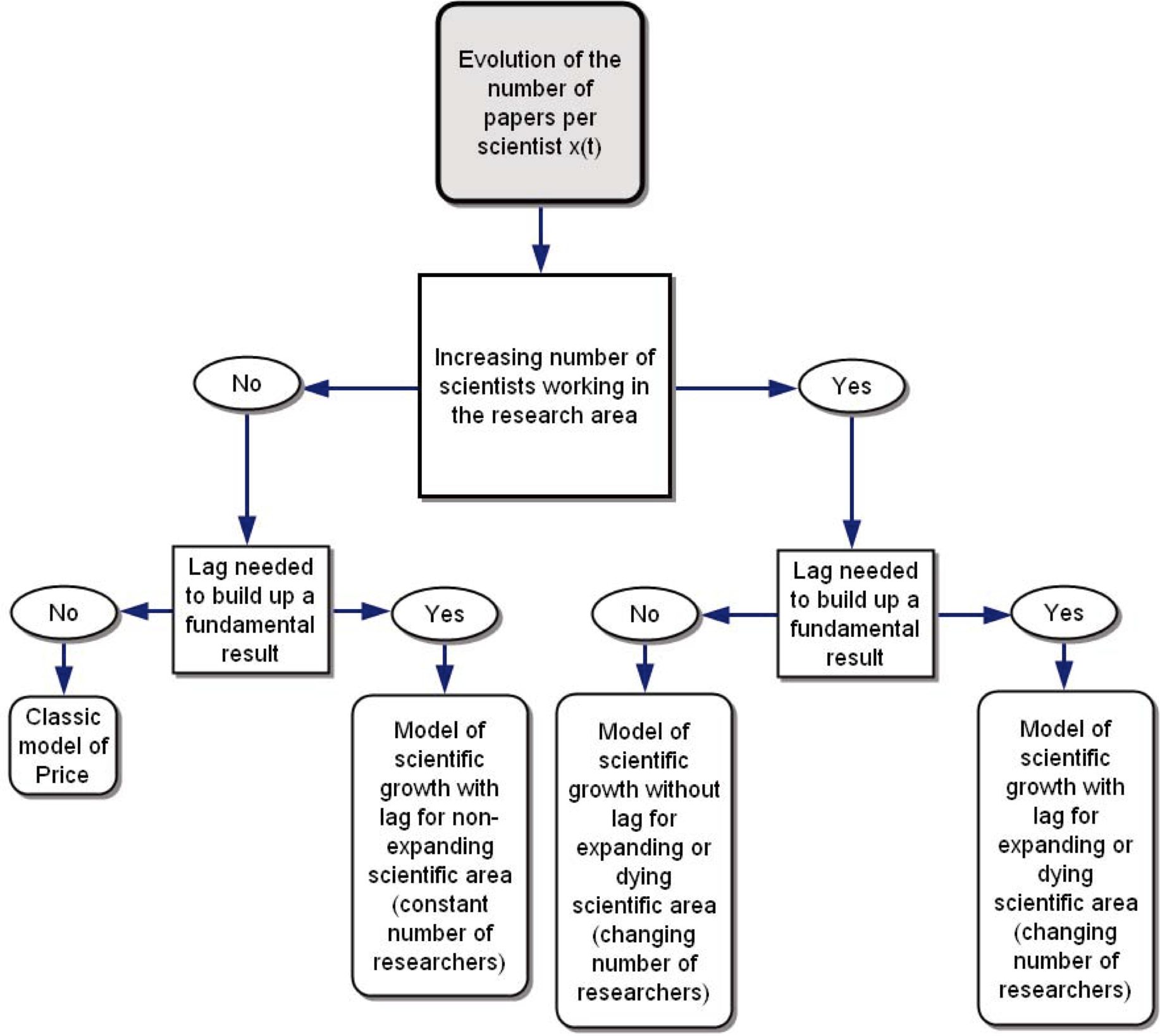}
\end{center}
\caption{Diagram of relationships between Price model and its modifications.
The presence of  time lags can lead to much complication in the  evolution  dynamics of
a scientific field. }
\end{figure}
The Price evolution model of scientific growth ignited intensive research  \cite{fern,kraw} 
(see Fig. 8). This model is in fact a dialectical addition to Kuhn's idea \cite{kuhn} 
about the revolutionary nature of science  processes: after some period of evolutionary 
growth, a scientific revolution occurs.
Price considered the exponential growth as a disease  that retards
the growth of stable science, producing narrower and less flexible specialists.
\begin{center}
{\fbox{\fbox{\parbox{13.0cm}{ \vskip0.5cm  \sf  FOR POLICY-MAKERS\\
Take-away box Nr.8 \\ An interesting result of the research of Price  can be read as follows:\\
{\bf if a government wants to double the usefulness of science, it has to multiply by 
about eight the gross number of workers and the total expenditure of manpower and 
national income}.
 \vskip0.5cm }}}}\\
\end{center} 
\par
The  unreserved  application of the Price model faces several difficulties:
\begin{itemize}
\item
many scientific products which seem to be new are not really new;
\item creativity and innovation can be confused \cite{plsek,amabile};
\item  creative papers with
new ideas and results have the same importance as  trivial duplications \cite{beck};
\item 
two things are omitted:
\begin{itemize}
\item
quality (whatever that means, but it is an economic notion) of research;
\item 
the cost or measure of complexity.
\end{itemize}
\end{itemize}
 In answer to this,  Price formulated the hypothesis that
one  should be studying only  the growth of  {\bf important}  discoveries, inventions, and scientific
laws, rather than both important and trivial things. In so doing, one might expect that any  of such studied growth will follow the
same pattern. 
\par
A generalized version of the Price model  for the growth
of a scientific field \cite{pol,price2} is based on the following assumptions: {\bf 
(a)} the growth is measured by the number of 
important publications appearing at a given time; {\bf (b)} the growth has a continuous
character, though a finite time period $T = {\rm const}$ is needed to build up a 
result of the fundamental character; {\bf (c)} the interactions between 
various scientific fields are neglected. If,  in addition, the number of scientists publishing 
results in this field is constant, then  the rate of scientific growth is proportional 
to the number of important publications at   time $t$ minus the time period $T$ 
required to build up a fundamental result.  The model equation is 
\begin{equation}\label{eqp1}
\frac{dx}{dt} = \alpha x(t - T),
\end{equation}
where $\alpha$ is a  constant. The initial condition $x(t) = \phi (t)$ is defined on the 
interval $[-T,0]$. 
\par
Let  the population of scientists be varying and consider the evolution of the average number 
of papers per scientist. In general, instead of the linear right-hand side Eq.(\ref{eqp1}), a 
non-linear model can be used:
\begin{equation}\label{eqp11}
\frac{dx}{dt}= f(x(t-T),x(t)), \end{equation}
where $f(t-T)$ is a homogeneous function of degree one. The simplest form of such a
function is a linear function. Let $n(t)$  represent  the rate of growth of the population of scientists
and write $L(t)=\exp[n(t)\;t]$. For simplicity, let the population of scientists
grow at the constant rate $n =\frac{1}{L} \frac{dL}{dt}$ and let $z = x/L$.  
Then   the evolution of the number of papers written by a scientist has the form
\begin{equation}\label{eqp7}
\frac{dz}{dt} = \alpha z(t-T) - n z(t).
\end{equation}
If $n=0$ and $T=0$,  the Price model of exponential growth is recovered. Eq.(\ref{eqp7}) is  
linear, but a cyclic behavior may appear because of the feedback between the delayed and 
non-delayed terms. 
\subsection{Models based on three or four populations. Discrete models.}
\par
{\bf (1) SIR (Susceptible-Infected-Removed) model (Fig.9)}\\
In 1927,   Kermack and   McKendrick \cite{kmk} created a model in which they 
considered a fixed population with only three compartments:  $S(t)$, the susceptibles;  $I(t)$, the infected;  $R(t)$, the
   recovered,  or removed.
\begin{figure}[h]
\begin{center}
\includegraphics[scale=0.6]{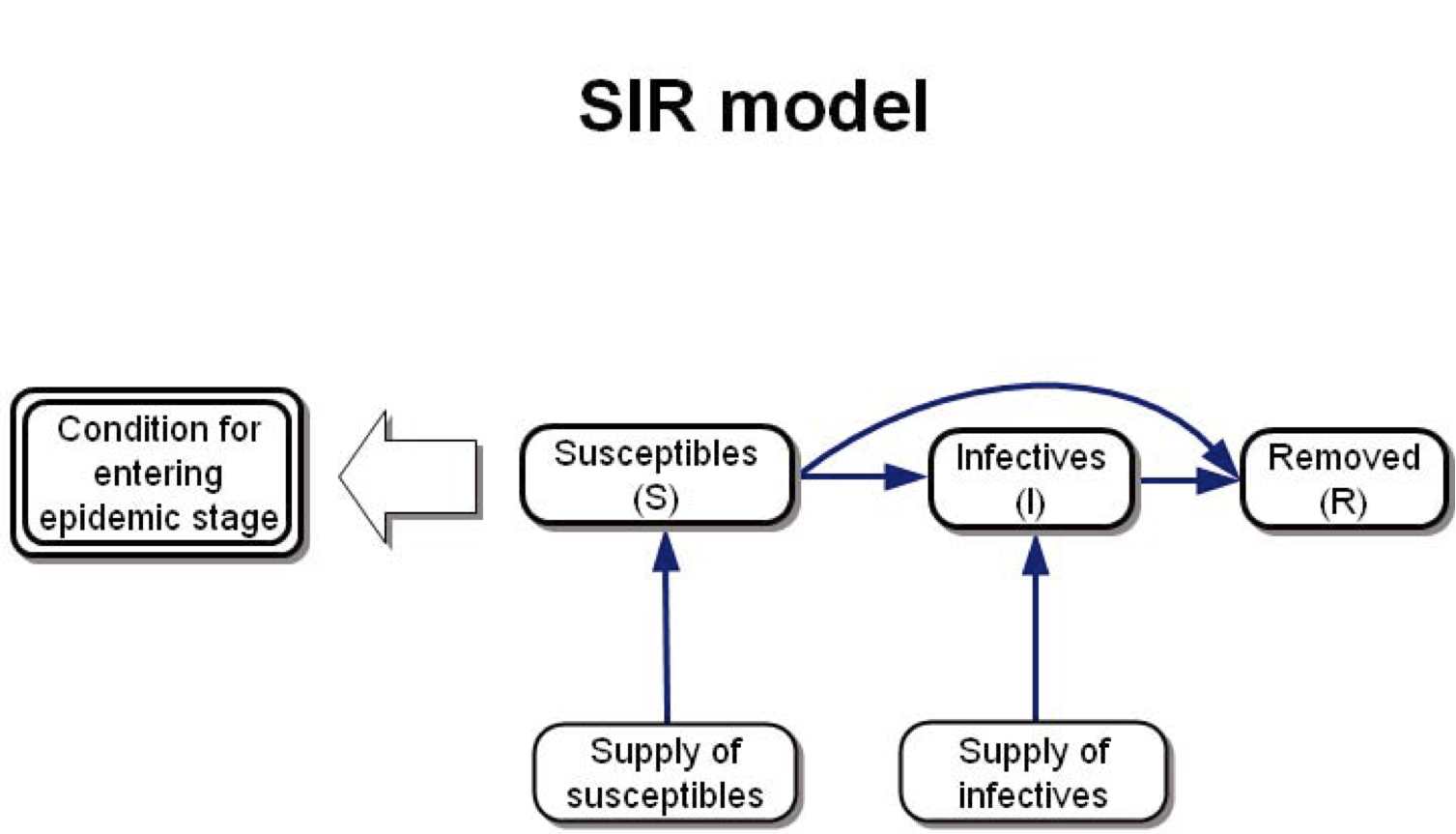}
\end{center}
\caption{SIR  (susceptibles $S$,  infectives $I$, recovered $R$) model of intellectual 
infection with influxes of susceptibles and infectives to the corresponding scientific ideas.}
\end{figure}
\par  
Following this idea,  Goffman and Newill \cite{gof2, gof1} considered 
the stages of fast growth of scientific research in a scientific field  as "intellectual 
epidemics" and developed  the corresponding 
 scientific research  epidemic stage based on  three  classes of population:  (i) the 
susceptibles $S$ who  can become infectives  when in contact with 
infectious material (the ideas); (ii) the infectives $I$ who host the infectious material; and   
(iii) the recovered $R$ who  are removed  from the epidemics for different 
reasons (Fig. 9).
\par
The epidemic stage is controlled by the system of  differential equations
\begin{eqnarray} \label{eqa1}
\frac{dS}{dt}=- \beta SI - \delta S + \mu , \\
\frac{dI}{dt}=\beta S I - \gamma I + \nu, \\
\frac{dR}{dt}= \delta S + \gamma I
\end{eqnarray}
where $\mu$ and $\nu$ are the rates at which the new supply of susceptibles and infectives 
enter  the population.  A necessary condition for the process to enter the epidemic state is
$\frac{dI}{dt} >0$. Then
\begin{equation}\label{eqa5}
S > \frac{\gamma - \nu/I}{\beta} = \rho
\end{equation}
is the threshold density of susceptibles, i.e., no epidemics can develop from time $t_0$
unless $S_0$, the number of susceptibles at that time, exceeds the threshold $\rho$: the 
epidemic state cannot be maintained over some time interval unless the number 
of susceptibles is larger than $\rho$
through that interval of time.  As $I$ increases, $\nu/I$ converges to $0$ and $\rho$
converges rapidly to $\gamma/\beta$. 
\par
In  \cite{gof2},  Goffman evaluated the rate of change of infectives $\Delta I /\Delta t$.
From the system  equations, it is difficult to determine $I(t)$. Yet
in the epidemic stage, the behaviour of $I(t)$ is exponential. For small
$t$ close to $t_0$, $I(t)$ can be expanded into a power series:
$I(t) = C_0 + C_1 t + C_2 t^2 + \dots C_n t^n+ \dots$
such that the approximate rate of $\Delta I/ \Delta t$ can be obtained.
On the basis of this rate and the raw data, the development and 
peak of some research activity can be predicted, -
under the assumption that the research is in an epidemic stage.
\par
{\bf (2) SEIR model for the spreading of scientific ideas (Fig. 10)}\\
The  SIR epidemic models can be further refined by introducing a  fourth class, $E$, i.e.,
persons exposed to the corresponding scientific ideas (Fig. 10). Such  models are
discussed in  \cite{bett2,bett1}; they belong to the class of 
so-called  SEIR epidemic models. One typical model goes  as follows
\begin{eqnarray}\label{eqb1}
\frac{dS}{dt} = \lambda N - \frac{\beta S I}{N};  \hskip.5cm 
\frac{dE}{dt}=\frac{\beta S I}{N} - \kappa E - \frac{\rho E I}{N};\\
\frac{dI}{dt} = \kappa E + \frac{\rho E I}{N} - \gamma I;  \hskip.5cm
\frac{dR}{dt}=\gamma I
\end{eqnarray}
where $S(t)$ is the size of the susceptible population at time $t$, $E(t)$ is the 
size of the exposed class, $I(t)$ is the size of the infected class. These individuals 
 have adopted the new scientific idea  in their publications. Finally, 
$R(t)$ is the size of the population of recovered scientists, i.e., those who no longer 
publish on the topic. The size of the entire population is: $N = S+E+I+R$. 
\begin{figure}[h]
\begin{center}
\includegraphics[scale=0.49]{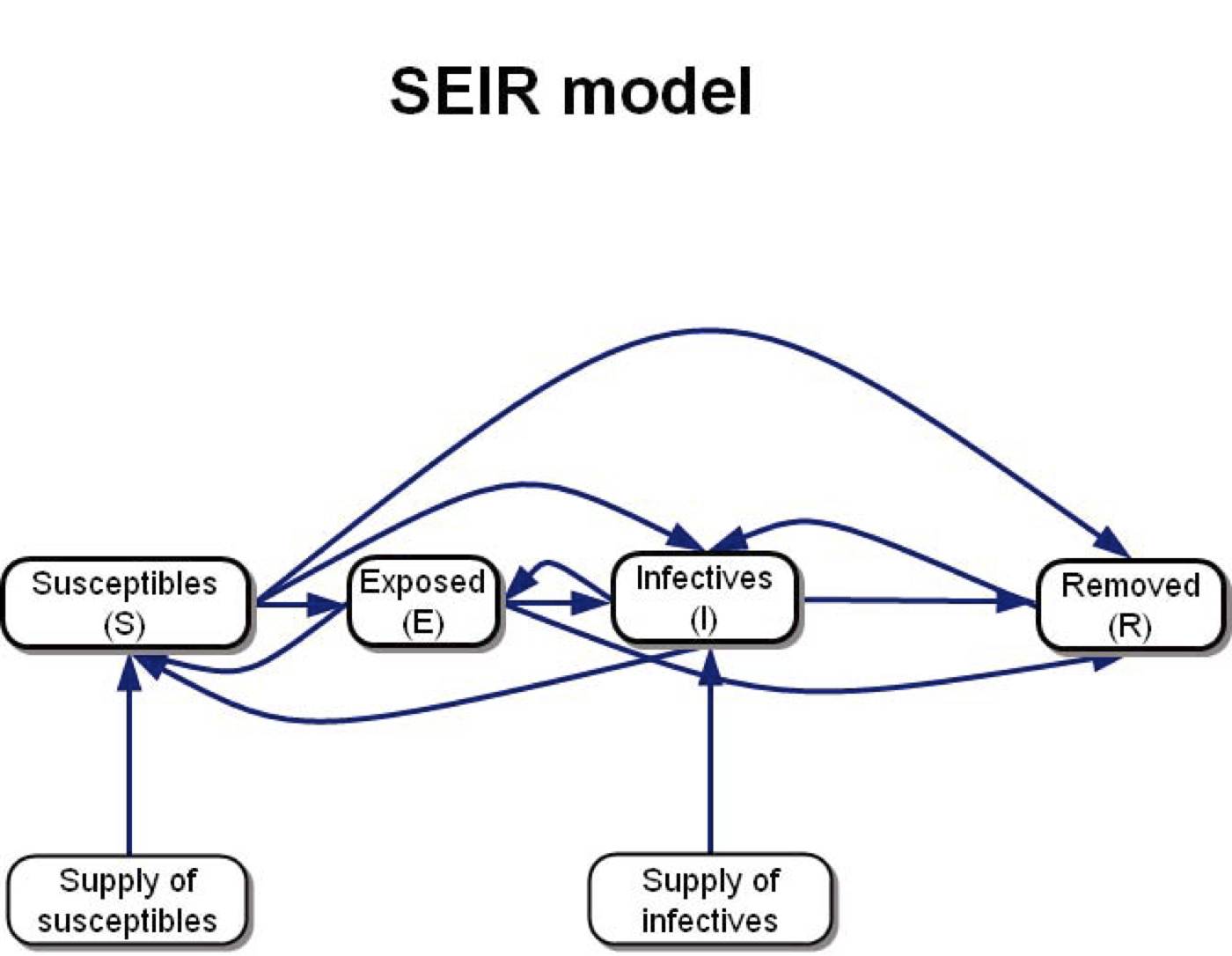}
\end{center}
\caption{SEIR model of intellectual infection with influxes of susceptibles
and infectives to the corresponding scientific ideas,  thus extending  the 
  SIR model by including a class of scientists exposed ($E$) to  the specific scientific ideas. }
\end{figure}
An exit  term is assumed to be very small, 
and because of this, $t$ is included in the recovered class.  $N$ grows
exponentially with rate $\lambda$. The parameters of the model  are: $\beta$, the probability 
and effectiveness of a contact with an adopter;  $1/\kappa$,
the standard latency time, (in other words, the average duration of time after 
one has been  exposed but before one includes the new
idea in one's own publication); $1/\gamma$, the duration of the infectious period, 
 thus how long one publishes
on the topic and  teaches others; $\rho$, the probability that an exposed person has
multiple effective contacts with other adopters.
\par
This simple model can incorporate a wide range of behaviors.
For many values of the parameters $\lambda$,$\beta$, $\kappa$, $\gamma$ and $\rho$, 
the infected class  grows  as a logistic curve. For large values of the contact 
rate $\beta$  or recruitment $\lambda$,  $I(t)$  grows  
nearly linearly, as indeed has been found empirically for some research fields \cite{bett2}.
\begin{center}
{\fbox{\fbox{\parbox{13.0cm}{ \vskip0.5cm \sf 
FOR POLICY-MAKERS\\
 Take away box Nr.9: \\Epidemic models are the best suited for describing  the {\bf expansion
stage}  of a process growth.   
 \vskip0.5cm }}}}\\
\end{center} 
\par
{\bf (3) SI discrete model  for the change in the number of  authors in a scientific field 
(Fig. 11) }\\
\begin{figure}[h]
\begin{center}
\includegraphics[scale=0.48]{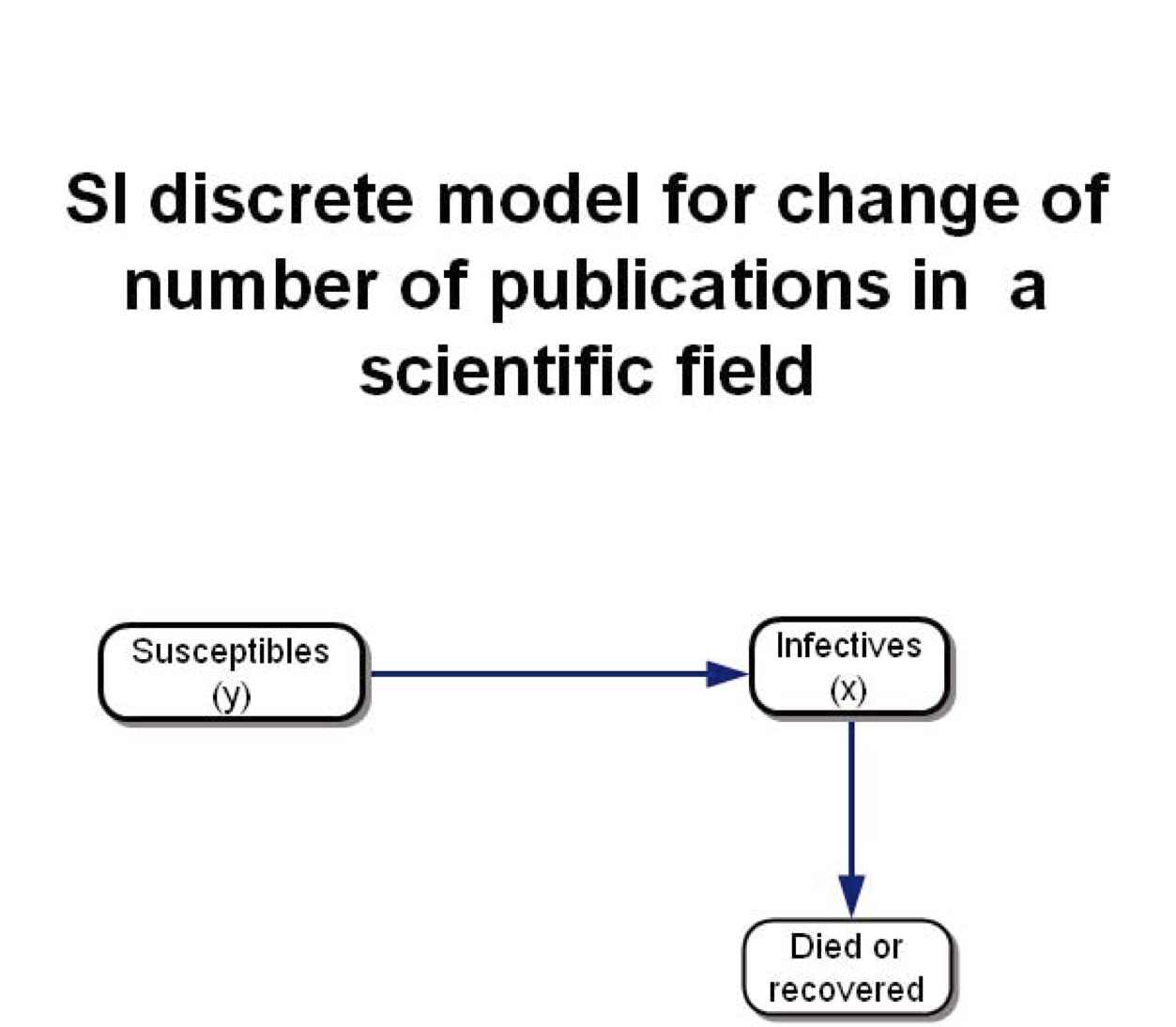}
\end{center}
\caption{Schema of a discrete SI evolution model of  the number of
authors of scientific papers. The model takes into account that several
scientists stop their work  in a scientific field; it can be due to
different reasons as for example  death or losing interest in particular questions.}
\end{figure}
With the  goal of predicting  the spreading out of scientific objects (such as theories
or methods),  Nowakowska \cite{n1} discussed several epidemic discrete 
models for predicting changes in the number of publications and authors  in a given 
scientific field. 
With respect to the publications, the main assumption of the models is that
the number of publications in the  next period of
time (say, one year) will depend: (i) on the number of papers which  recently
appeared, and (ii) on the degree at which the subject has been
exhausted. The numbers of publications appearing in successive periods of time should 
first increase, then would reach a maximum, and  as the problem
becomes more and more exhausted, the number of publications would decrease.
\par
Let it be  assumed (Fig. 11) that  if at a certain moment $t$ the  epidemics state  is ($x_t, 
y_t$) ($x_t$ is the number of infectives (authors who write papers on the corresponding 
research problems) , $y_t$ is the number of susceptibles), then for a 
sufficiently short time interval $\Delta t$, one may expect that the number of infectives 
$x_{t+\Delta t}$ will be equal to $x_t - a x_t \Delta t + b x_t y_t \Delta t$,
while the number of susceptibles $y_{t + \Delta t}$ will be equal 
to $y_t - b x_t y_t \Delta t$; $a$ and $b$ being appropriate constants. Let
the expected number  
of individuals who either die or recover, during the interval ($t, t + \Delta t$), be $a x_t 
\Delta t$, and let  $b x_t y_t \Delta t$  be the expected number of new infections.
The equations of this model are:
\begin{eqnarray}\label{eq_nov1}
x_{t +\Delta t} &=& a x_t -a x_t \Delta t + b x_t y_t \Delta t  \\
y_{t+\Delta t} &=& y_t - b x_t y_t \Delta t.
\end{eqnarray}
Note here that such discrete models are useful for the analysis of realistic  situations 
where the values of the quantities are available at selected moments  (every month, every 
year, etc.).
\par
{\bf (4)  Daley  discrete model for the population of papers (Fig. 12)}\\
\begin{figure}[h]
\begin{center}
\includegraphics[scale=0.5]{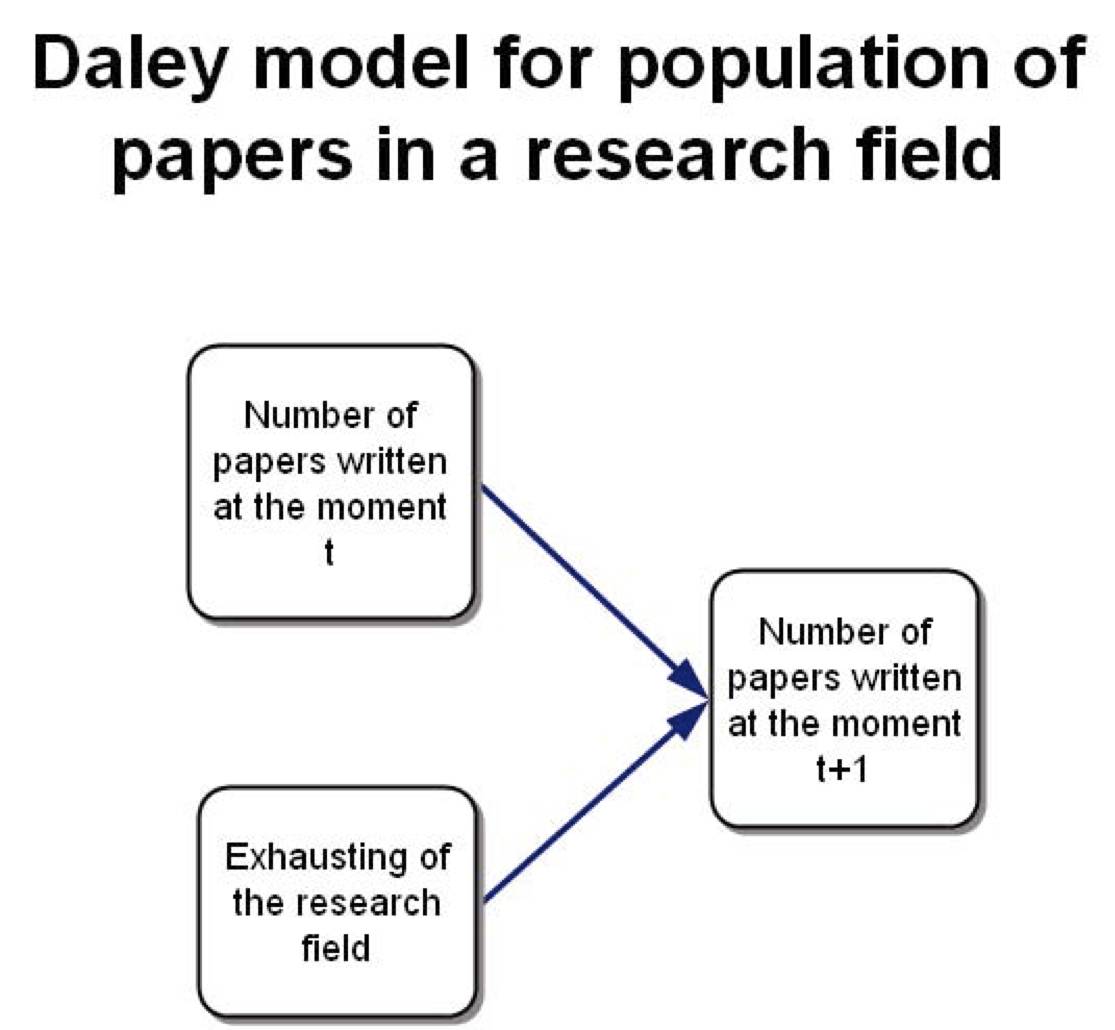}
\end{center}
\caption{Daley model for evolution of population of papers on problems
in a scientific field. The exhausting of the scientific field is taken into
account.}
\end{figure}
Daley \cite{daley} investigated  the spread of news as follows: individuals who have not
heard the news are susceptible and those who heard the news are
infective. Recovery is not possible, as it is assumed that the individuals
have perfect memory and never forget. The Daley model  can be applied also to the
population of papers \cite{n1} (see Fig. 12). For $\Delta t = 1$ (year), the 
Daley model equation  reads
\begin{equation}\label{eq_nov3}
x_{t+1} = b x_t  \left( N - \sum_{i=1}^t x_i \right)
\end{equation}
where $x_1$, $x_2$ .... are the numbers of papers on the subject
which appear in successive periods of time, $b$ and $N$ being 
parameters. The expected number $x_{t+1}$ of papers in year 
$t + 1$ is proportional to the number $x_t$ of papers which appeared in year $t$, and to the
number $N - x_1 - x_2 \dots - x_t  =  N- \sum_{i=1}^t x_i$. $N$ is the number of papers which have to appear in order to exhaust the problem: the problem
under consideration may be partitioned into $N$ sub-problems, such that
solving any of them is worth a separate publication;  these subproblems
are solved successively by the scientists.
The  $b$ and $N$ parameters may be estimated 
by the method of least squares, e.g.  from a given empirical histogram. A parameter 
characterizing the initial growth dynamics   in the  number of publications can also be 
introduced: $\tau =b N$. Therefore, Eq.(\ref{eq_nov3}) can be used for short-time prediction, even when the corresponding research field is 
in the epidemic stage of its evolution.
\par
{\bf (5)  Discrete model coupling  the populations of scientists and papers (Fig. 13)}\\
\begin{figure}[h]
\begin{center}
\includegraphics[scale=0.32]{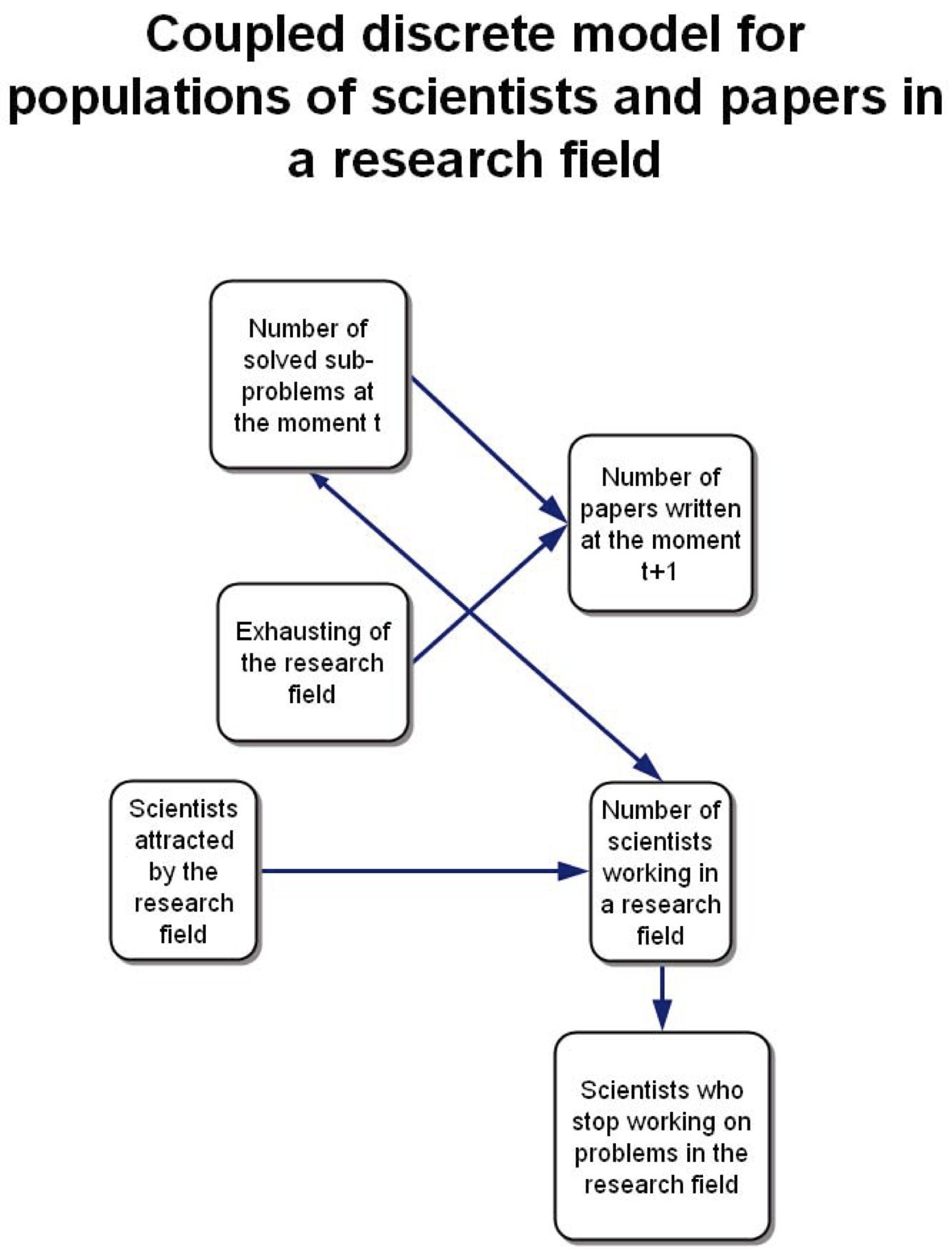}
\end{center}
\caption{ Discrete model for  the joint evolution of populations of
scientists and papers. The attractiveness of the field, the exhaustion
of the field, and the possibility for declining  interest for working in
the scientific field are taken into account through adequate rate parameters.}
\end{figure}
A discrete model coupling the populations of scientists and papers can be considered (Fig. 13); 
it depends on four parameters: $N$, $a$, $b$ and $c$. $N$ as above denotes the number
of sub-problems of the given problem;  $a$ is the probability that a  scientist working on
the subject in a given year abandons  
research on the subject for whatever reasons;  $b$  is the probability of obtaining a
solution to a given subproblem by one scientist during one year of research;  
$c$ denotes the coefficient of attractiveness of the subject. The
basic variables of the model are:
$u_t$, the number of scientists working on the subject in year $t$, and 
$x_t$, the number of publications on the subject which appear in year $t$.
\par
The model equations are   
\begin{eqnarray}\label{eq_nov4u}
u_{t+1} &=& (1 - a) u_t + c x_t  \\
x_{t+1} &=& [1 -(1-b)^{u_t}]\left( N - \sum_{i=1}^{t} x_i \right).
\end{eqnarray}
The equation for the number $u_{t+1}$ of scientists working on the
subject in year $t + 1$ tells   that in year $t + 1$, the expected
number of scientists working on the subject will be
 the number of scientists working on the subject in year $t$, $u_t$,
minus  the expected number of scientists who stopped working on the
subject,  $a u_t$, plus the expected number of scientists,  $c x_t$, who became attracted to 
the problem by reading papers which appeared in year $t$.
The equation expressing the number of publications in year $t + 1$ tells us that  
$x_{t+ 1}$  equals   the number of subproblems that were solved in the year $t$.
The probability that a given subproblem will be solved in year $t$
by a given scientist equals $b$. Then  the probability of the opposite event,
i.e.   a given scientist  will not solve a particular problem,
equals $1- b$. As there are $u_t$ scientists working on the subject in year $t$, the 
probability that a given subproblem will not be solved by any of them is
$(1 - b)^{u_t}$. Consequently, the probability that a given subproblem will be solved in
year $t$ (by any of the $u_t$ scientists working on the subject) is equal to  $1 - (1 - 
b)^{u_t}$.
Next, in year $t$ there remained $N - \sum_{i=1}^{t} x_i$
subproblems to be solved.  The expected number of subproblems
solved in year $t$ is equal to the product  which gives the right-hand side of   Eq.(28).  
\par
N.B. 
It is assumed, that the waiting time for publishing of the paper is one year. A more 
realistic picture would be  to  assume  that the unit of time is not one year, 
but two years, or  that the publication has some other time delay.
\begin{center}
{\fbox{\fbox{\parbox{13.0cm}{ \vskip0.5cm \sf 
FOR POLICY-MAKERS\\
 Take away box Nr.10: \\ In many cases, the data  is  available as one value per
week, or one value per month, or one value per three months, etc. 
For modeling and subsequent short-range
forecasting, so-called   discrete (time) models are thus very appropriate.  
 \vskip0.5cm }}}}\\
\end{center} 
\subsection{ Continuous models of the joint evolution of scientific sub-systems} 
\par
{\bf (1) Coupled continuous model for the populations of scientists and papers:  
Goffman-Newill model }\\
\begin{figure}[h]
\begin{center}
\includegraphics[scale=0.3]{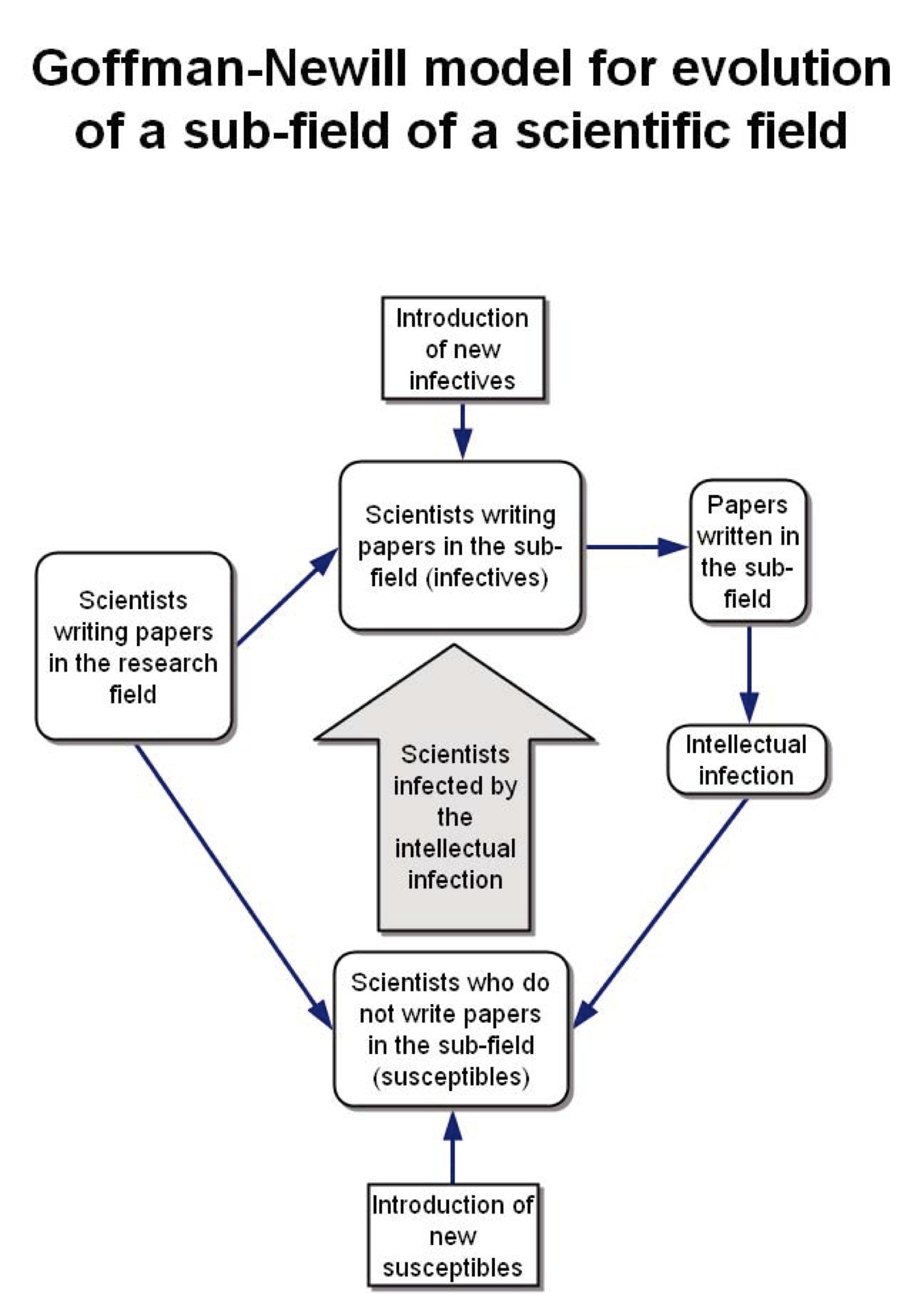}
\end{center}
\caption{Schema of Goffman-Newill model for the evolution of  a scientific
field. Scientists are attracted to a sub-field after being intellectually
infected  by papers from the sub-field. }
\end{figure}
The Goffman-Newill model \cite{gof1} (Fig. 14)  is based on the idea 
that the spreading process 
within a population  can be studied on the basis of the literature 
produced by the members of that population. There is a transfer of 
infectious materials (ideas) between humans by means of an  intermediate host (a written 
article).
Let   a scientific field be $F$ and $SF$ a sub-field  of $F$. Let the number of scientists
writing papers in the field $F$ at   $t_0$ be $N_0$ and the number of scientists
writing papers in $SF$ at   $t_0$ (the number of infectives) be  $I_0$.  Thus, $S_0 = N_0 - I_0$ is the number of susceptibles; there is no removal at $t_0$, but there 
is removal $R(t)$ at later times $t$. The number of papers produced 
on $F$ at $t_0$ is $N_0'$ and the number of papers produced in 
  $SF$ at this time is $I_0'$. The process of intellectual infection is as follows:  
{\bf (a)} a member of $F$ is infected by a paper from $I'$; 
{\bf (b)} after some latency period,
this infected member produces 'infected' papers in $N'$, i.e. the infected member produces
a paper in the subfield $SF$   citing a paper from $I'$; 
{\bf (c)}
this 'infected' paper may infect other scientists from $F$ and its sub-fields, such that
the intellectual infection spreads from $SF$ to the other sub-fields of $F$.
\par
Let $\beta$ be the rate at which the susceptibles from  class $S$
become 'intellectually infected' from   class $I$. Let $\beta'$ be the rate at 
which the papers in $SF$ are cited  
by members of $N$ who are producing papers in $SF$. As the infection process develops, some
susceptibles and infectives are removed, i.e. some scientists are no longer active,
and some papers are not cited anymore. Let 
$\gamma$ and $\gamma'$ be the rates of removal of infectives from the populations $I$ and
$I'$  respectively, and $\delta$ and $\delta'$ be the rates of removal from the populations of
susceptibles $S$ and $S'$. In addition, there can be a supply of infectives and susceptibles 
in  $N$ and $N'$. Let the rates of introduction of new susceptibles 
be $\mu$ and $\mu'$, i.e. the rates at which the new authors and new papers are introduced 
in $F$, and  let the rates of introduction of new infectives be $\upsilon$ and
$\upsilon'$, i.e. the rates at which new authors and new papers are introduced in $SF$. In 
addition, within a short time interval  a susceptible can remain susceptible 
or can become an infective or  be removed; the infective can remain an infective or can become
a removal; and the removal remains a removed. The immunes remain immune and do not return to  
the population of susceptibles.  If, in addition, the populations are  homogeneously mixed,  
the system of model equations  reads   
\begin{eqnarray}\label{gof1}
\frac{dS}{dt} &=& - \beta S I' - \delta S + \mu ; \hskip.5cm
\frac{dI}{dt} = \beta S I' - \gamma I + \upsilon \\
\frac{dR}{dt} &=& \gamma I + \delta S; \hskip.5cm
\frac{dS'}{dt} = - \beta' S' I - \delta S' + \mu' \\
\frac{dI'}{dt} &=& \beta' S' I - \gamma' I' + \upsilon'; \hskip.5cm
\frac{dR'}{dt} = \gamma' I' + \delta' S'
\end{eqnarray}
The conditions for development of an epidemic are as follows. If   as an initial
condition at $t_0$, a single infective is introduced into the populations $N_0$ and
$N'_0$, then for an epidemic to develop, the change of the number of infectives must be 
positive  in  both populations. Then, for
$
\rho = \frac{\gamma - \upsilon}{\beta}$ and $ \rho' = \frac{\gamma' - \upsilon'}{\beta'},
$
the threshold for the epidemic arises from  the conditions $\beta S I' > \gamma I - 
\upsilon$ and
$\beta' S' I' > \gamma' I' - \upsilon'$, such that the threshold is
\begin{equation}\label{gof_thres}
S_0 S'_0 > \rho \rho'.
\end{equation}
The development of epidemics is given by the equation $\frac{dI}{dt} = D(t)$.
The peaks of the epidemic occur at  time points where $\frac{d^2 I}{dt^2} =0$, while  the
epidemic's size is given by $I(t \to \infty)$.
\par
{\bf (2) Bruckner-Ebeling-Scharnhorst model for the growth of $n$ subfields in a
scientific field}\\
\begin{figure}[h]
\begin{center}
\includegraphics[scale=0.35]{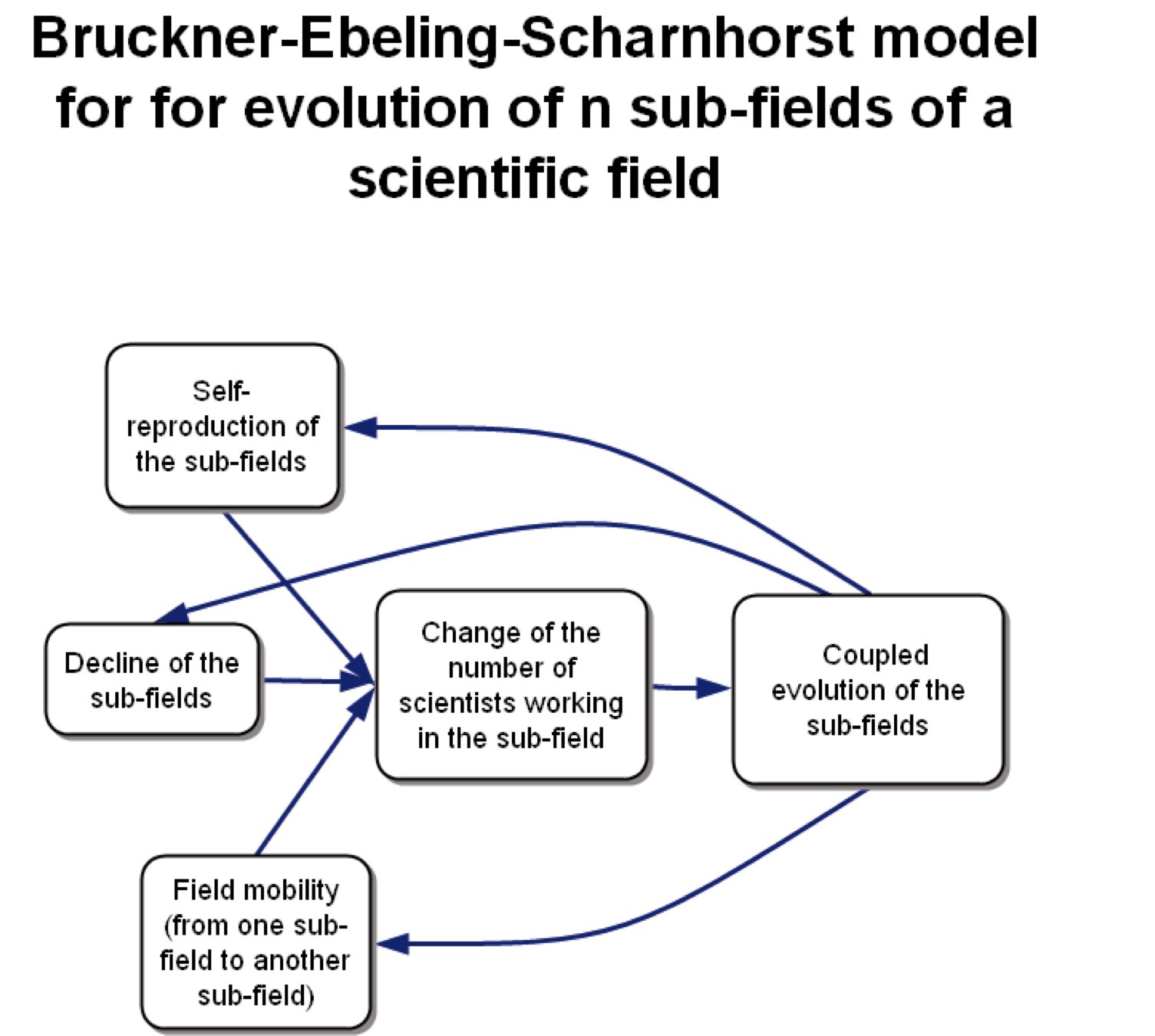}
\end{center}
\caption{Schema of Bruckner-Ebeling-Scharnhorst  model of evolution of
$n$ scientific sub-fields. Self-reproduction and decline of subfields
as well as field mobility are taken into account.}
\end{figure}
The evolution of growth processes in a system of scientific fields can be modeled by  
complex continuous evolution models. One of them,
the Bruckner-Ebeling-Scharnhorst approach \cite{andrea1} (Fig. 15),   is closely related to 
several generalizations of  Eigen's theory of prebiotic evolution  and is briefly discussed 
here (see also \cite{ax22}). In 1912, Lotka \cite{lotka} published the idea of describing biological 
epidemic processes, like malaria, as well as chemical oscillations, with the help 
of a set of differential equations. These equations, known as  Lotka-Volterra 
equations \cite{lotka1,volterra}, are  used to  describe a coupled growth process 
of populations. However,  they do not reflect several essential properties of 
evolutionary processes such as the creation of new structural elements. 
Because of this, one has to consider a more general set of equations for the change in the 
number $x_i$ of the scientists from the $i$-th scientific subfield (a Fisher-Eigen-Schuster 
kind of model), i.e.,
\begin{eqnarray}\label{eqf2}
\frac{dx_i}{dt} = (A_i - D_i)x_i + \sum_{j=1; j\ne i}^n (A_{ij} x_j - A_{ji} x_i) +
\sum_{j=1; j\ne i}^{n} B_{ij} x_i x_j - k_0 x_i, \nonumber \\
i,j = 1,\dots,n.
\end{eqnarray}
The model based on Eq.(\ref{eqf2}) describes the coupled growth of $n$ 
subfields, of a scientific discipline. Three fundamental processes of evolution 
 are  included in Eq.(\ref{eqf2}) : {\bf (a)}
self-reproduction: students and young scientists join the  field  
and start working on corresponding problems. Their choice is influenced mainly
by the education process as well as by individual interests and by existing
scientific schools; {\bf (b)}
decline: scientists are active in science for a limited number of years. For 
different reasons (for  example, retirement)  they stop working and leave the 
 system; {\bf (c)}
field mobility:  individuals turn to other fields of research for various reasons or  maybe 
open up new ones themselves. 
\par
 The reasoning to obtain Eq.(\ref{eqf2})  goes as follows. The general form of
the law for growth of the $i$-th subfield is supposed to be 
\begin{equation}\label{eqf31}
\frac{d x_i}{dt} =  f_i 
(\vec{x}),  \hskip.5cm \vec{x}=(x_1 , ...,x_n). \end{equation}
 By separation, $f_i = w_i x_i$,  one obtains the 
replicator equation
\begin{equation}\label{eqf32}
\frac{d x_i}{dt} = w_i x_i, \hskip.5cm i=1,2,\dots,n.
\end{equation}
 Notice that   when $w_i={\rm const}$,  the  fields are  uncoupled, i.e.,  there is an
  exponential growth in science.
Otherwise, $w_i$ itself is a function of $x$ and of various parameters, but can be separated 
into three terms according to the above  model assumptions , i.e.,
\begin{equation}\label{eqf4}
w_i = A_i - D_i + \sum_{j=1,j\ne i}^n \left(A_{ij} \frac{x_j}{x_i} - A_{ij} \right).
\end{equation}
Eq.(\ref{eqf2})  is thus  obtained from Eq.(\ref{eqf32}) and Eq.(\ref{eqf4})  for $B_{ij}=0$, 
$k_0 =0$.
To adapt this model to real growth processes,  it  can be assumed  
that the coefficients $A_i$, $D_i$, and $A_{ij}$ themselves are functions of $x_i$:
\begin{eqnarray}\label{eqf5}
A_i = A_i^0 + A_i^1 x_i + \dots; \hskip.15cm  D_i = D_i^0 + D_i^1 x_i + \dots; \hskip.15cm
A_{ij} = A_{ij}^0 + A_{ij}^1 x_j + \dots
\end{eqnarray}
Each of the three fundamental processes of change is represented in Eq.(\ref{eqf2}) 
with a linear and a quadratic term only. For example, the terms $A_i^1$ and $D_i^1$ account for 
cooperative effects in 
self-reproduction and decline processes  respectively, while $D_i^0$  accounts for a  
decline,  because of  aging. The contributions $A_{ij}^0$  assume a linear type of 
field mobility behavior for scientists analogous to a diffusion process.  On the other hand,  
the terms $A_{ij}^1$ represent a directed process of exchange of scientists between fields. 
The best way to obtain these parameters  is to estimate them  for specific  data bases using  
the method of least squares. 
\begin{center}
{\fbox{\fbox{\parbox{13.0cm}{ \vskip0.5cm \sf 
FOR POLICY-MAKERS\\
 Take away box Nr.11: \\The Bruckner-Ebeling-Scharnhorst model does not belong to the class 
of epidemic models which are best applicable only for describing the expansion stage of a 
process. \\The 
Bruckner-Ebeling-Scharnhorst model is an evolution model: it describes all stages of the 
evolution  of a system. 
 \vskip0.5cm }}}}\\
\end{center} 
\section{Small-size scientific and technological systems. Stochastic models (Fig. 16)}
\begin{figure}[h]
\begin{center}
\includegraphics[scale=0.7]{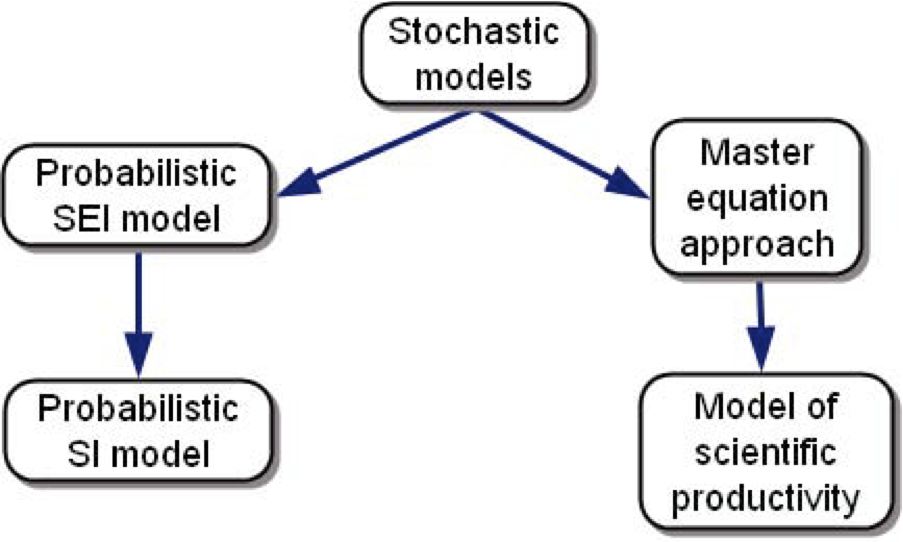}
\end{center}
\caption{Hierarchy of stochastic models discussed in this chapter.}
\end{figure}
The movement of large bodies in mechanics is governed by deterministic laws. When the body 
contains a small number of molecules and atoms, stochastic effects such as the
Brownian motion become important. In the area of scientific systems,
the fluctuations become very important when the number of scientists in a certain research 
subfield is small. This is typical for new research fields with only a few researching 
scientists.
\par
Several examples of stochastic models for the description of  the diffusion of ideas or 
technology and the evolution of science are:   {\bf (a)} the model of evolution of scientific 
disciplines with an example  pertaining to the case of  elementary particles physics  
\cite{kot};   {\bf (b)} stochastic models for the aging of scientific literature \cite{g2};   
{\bf (c)} stochastic models of the Hirsch index \cite{bur07} and of instabilities in
evolutionary systems \cite{br89};   {\bf (d)} models  of implementation of technological 
innovations \cite{br96},
etc. \cite{braun85}.  In the following,   see Fig. 16,  two probabilistic and two stochastic
 models  are discussed. Some attention is devoted to the master equation approach as well.
\subsection{Probabilistic SI and SEI  models}
Epidemiological models  of  differential-equation-based compartmental
type have been found to be limited in their capacity to capture heterogeneities at the 
individual level and in the interaction between individual epidemiological units \cite{ch04}.
This is one of the reasons to switch from   models  in which the
number of individuals  are in given  known states to  models involving  probabilities.
One such  model   \cite{kiss} captures the diffusion of topics over a network of connections 
between scientific disciplines, as assigned by the ISI Web of Science's classification in 
terms of Subject Categories (SCs). Each SC is considered as a node of a network along with
all its directed and weighted connections to other nodes or SCs \cite{kiss,kiss1}. As with
epidemic models, nodes can be characterized in a medical way. SCs that are  susceptible ($S$) 
are either not aware of a particular research topic or, if aware, may not be ready to adopt 
it. Incubating SCs ($E$) are those 
that are aware of a certain topic and have moved to do some research on problems connected with
this topic.  Infected SCs  ($I$) are actively working and publishing in a 
particular research topic.
\par
Two probabilistic models, i.e., (i)  the Susceptible-Exposed-Infected (SEI) model (Fig. 17) 
and (ii) 
a simpler Susceptible-Infected (SI) model (Fig. 18),  are thereby only  discussed.
\par
{\bf (1)   Susceptible-Exposed-Infected (SEI)   model}\\
\begin{figure}[h]
\begin{center}
\includegraphics[scale=0.44]{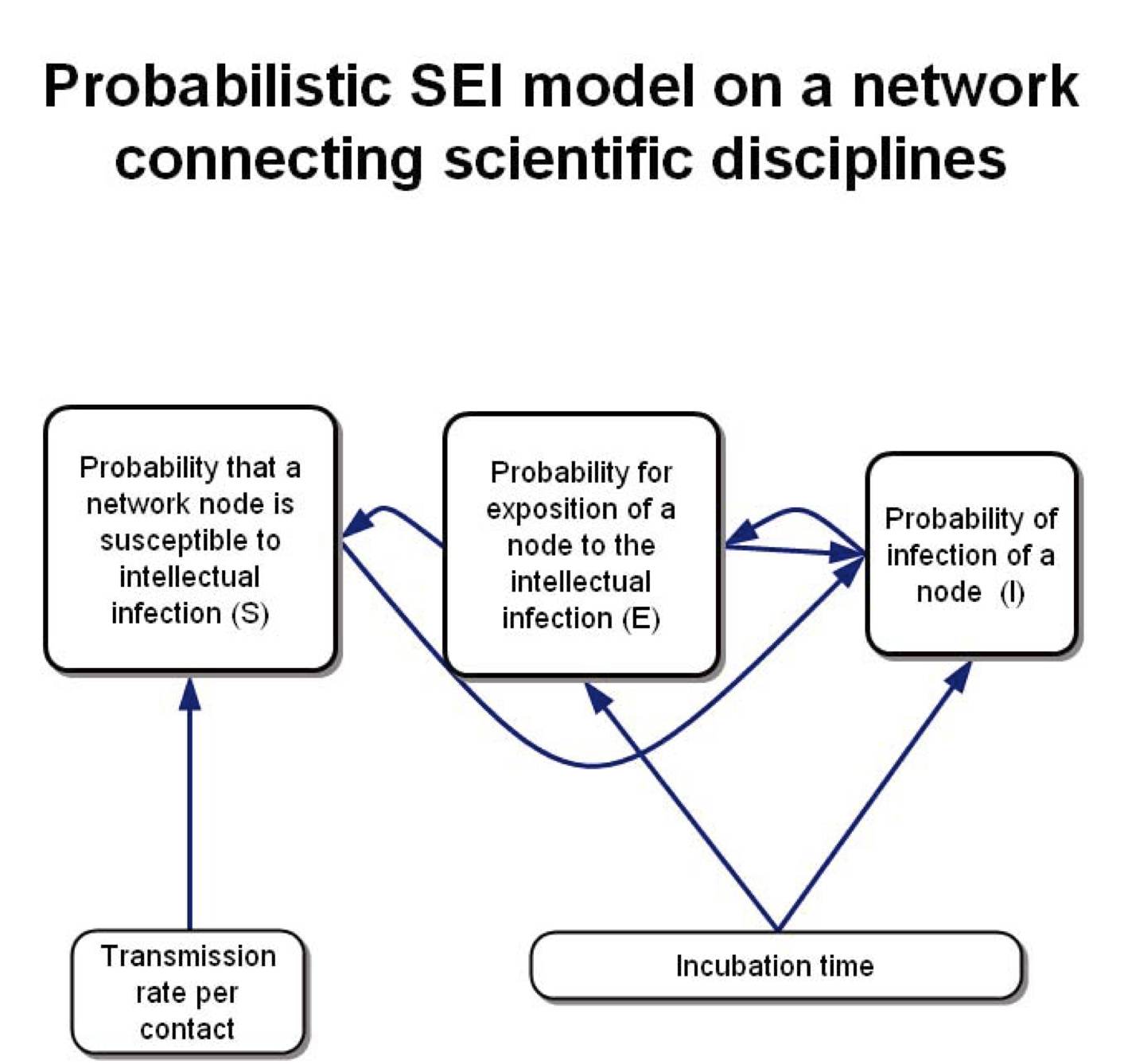}
\end{center}
\caption{Schema of the probabilistic SEI model for epidemics in a network 
connecting scientific disciplines.}
\end{figure}
The SEI model equations for the evolution of the node state probabilities are given by \cite
{kiss}:
\begin{equation}\label{eqc1}
\frac{dS_i(t)}{dt}= -\sum_j A_{ji}I_j(t)S_i(t), \hskip.5cm \end{equation}
\begin{equation}\label{eqc2} \frac{dE_i(t)}{dt}= \sum_j A_{ji}I_j(t)S_i(t)- \gamma E_i(t),
\end{equation}
\begin{equation}\label{eqc3}
\frac{dI_i(t)}{dt} = \gamma E_i(t),
\end{equation}
where $ 0 \le I_i (t) \le 1$ denotes the probability of node $i$ being infected at 
time $t$ (likewise for $S_i(t)$ and $E_i(t)$). The directed and weighted 
contact network is represented by $A_{ij} = r \Gamma_{ij}$ with
$\Gamma_{ij}$ = $(w_{ij})_{i,j=1, . . ., N}$ denoting the adjacency matrix that includes 
 weighted links; $r$ is the transmission rate per contact and $1/\gamma$ is
the average incubation or latent period. 
\par
This set of equations states that an increase  in the probability $E_i$ of a node $i$ being 
exposed  to  an infection is directly proportional to the probability $S_i$ of node $i$ being 
susceptible and  the probability $I_j$ of neighbouring nodes $j$ being infected. The number 
of such contacts and the per-contact  rate of transmission are incorporated in $A_{ij}$.  
Likewise, $E_i$ decreases if exposed/infected nodes become infected after an average 
incubation time 
$1/\gamma$. The number of   infected SCs at time  $t$, according to the model, 
can be estimated as $I(t) = \sum_i I_i(t)$.  Since $S_i (t) + E_i (t) + I_i (t) = 1$, 
for each $t > 0$, Eqs. (\ref{eqc1}) - (\ref{eqc3}) are readily understood, in view of Eq.(\ref{eqc2}).
\par
{\bf (2) Susceptible-Infected (SI) model}\\
\begin{figure}[h]
\begin{center}
\includegraphics[scale=0.4]{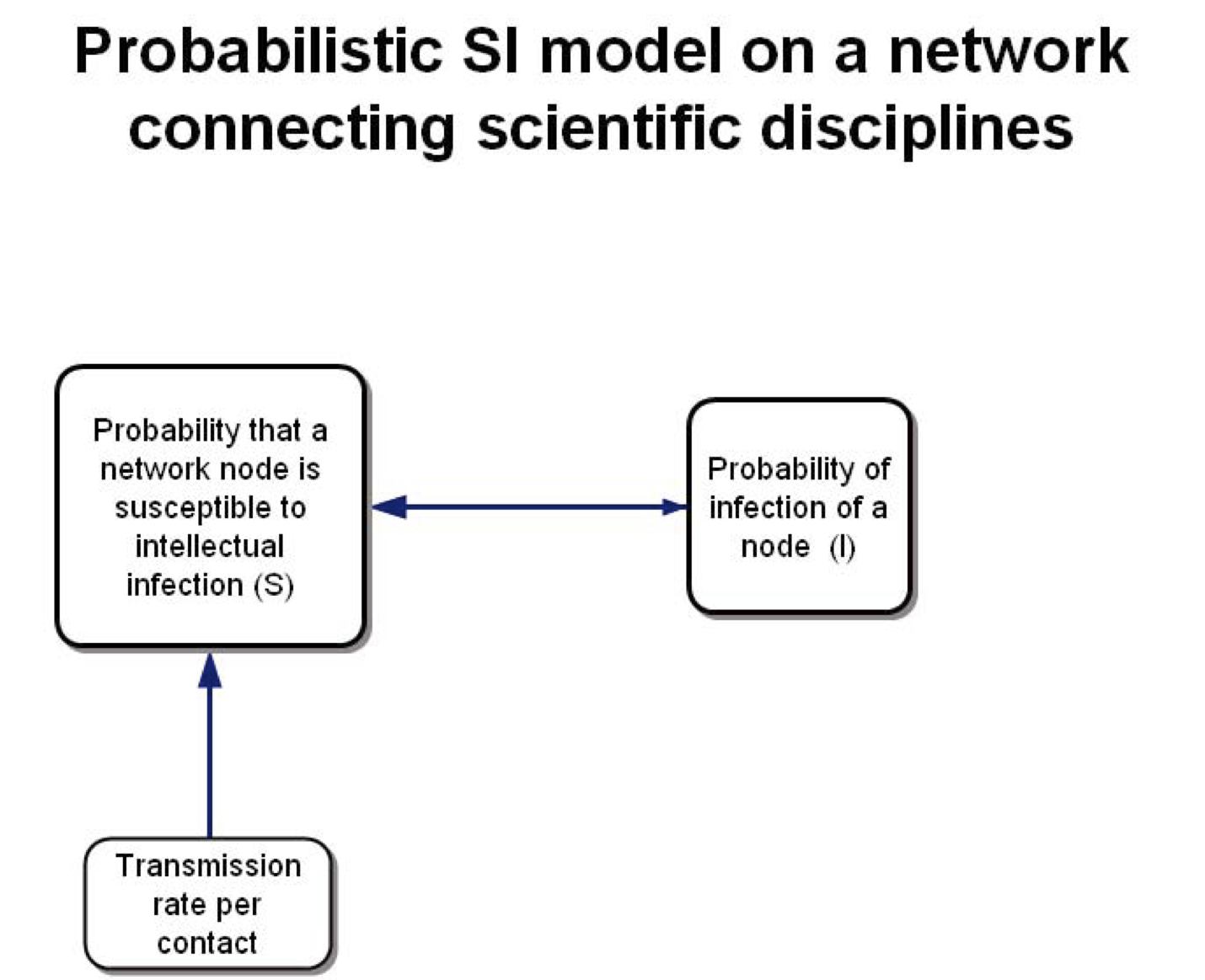}
\end{center}
\caption{Schema of the probabilistic SI model for epidemics in a network 
connecting scientific disciplines.}
\end{figure}
The above SEI model   can be simplified to  an SI model when the possibility 
of an exposed period is excluded, i.e,. if $\frac{dE_i(t)}{dt} = 0$. The equations for this 
simpler SEI model  are reduced to
\begin{equation}\label{eqc4}
\frac{dS_i(t)}{dt} = - \sum_j  A_{ji} I_j(t) S_i(t); \hskip.5cm
\frac{dI_i(t)}{dt} = \sum_j A_{ji} I_j(t) S_i(t),
\end{equation}
where the probability $I_i$ of a node $i$ being infected and infectious only 
depends on the probability $S_i$ of the node $i$ being susceptible.  The comparison of  both 
models with  available data shows  \cite{kiss} that
while the agreement at the population level is usually much better for the SEI model, for the 
same pair of parameters, the agreement at the individual level is better  when the simpler 
SI model is used. 
\subsection{Master equation approach} 
\par
{\bf (1) Stochastic evolution model with self-reproduction, decline, and
field mobility} \\
\begin{figure}[ht]
\begin{center}
\includegraphics[scale=0.38]{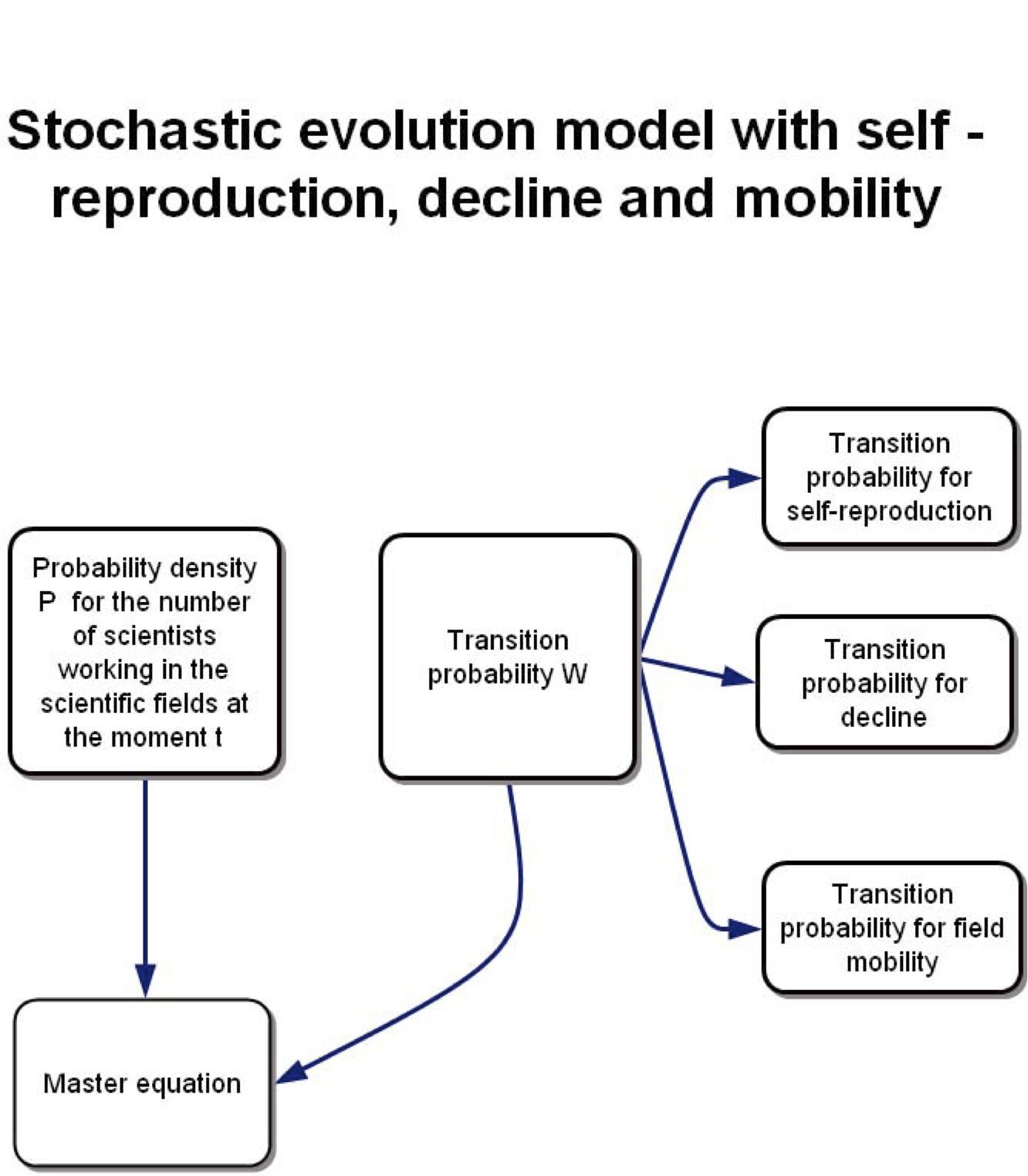}
\end{center}
\caption{Schema of the master equation model of evolution of scientific fields in
presence of self-reproduction, decline, and field mobility.}
\end{figure}
There exists a  high correlation between field mobility processes and the emergence of new 
fields \cite{andrea1}. This can be accounted for by a stochastic model (see Fig. 19), 
in which the system 
at  time $t$ is characterized by a set of integers $N_1$ , $N_2$, ..., $N_i$, ..., 
$N_n$, with $N_i$ being, e.g., the number of scientists working in the subfield $i$, 
which is considered now as a stochastic variable. The three fundamental types of scientific 
change mentioned in the discussion of the Bruckner-Ebeling-Scharnhorst model (see above)
here correspond to three elementary stochastic processes with three different
transition probabilities: \begin{itemize} \item {\bf (a)}
For self-reproduction, the transition probability is given by  \\
$
W(N_i + 1 \mid N_i) = A_i^0 N_i = A_i^0 N_i + A_i^1 N_i (N_i -1)
$;  \item 
{\bf (b)}
The transition probability for decline is  \\
$
W(N_i -1 \mid N_i) = D_i^0 N_i + D_i^1 N_i (N_i-1)
$;  \item 
{\bf (c)}
The transition probability for field mobility is  \\
$
W(N_i +1, N_j -1 \mid N_i N_j) = A_{ij}^0 N_j + A_{ij}^1 N_i N_j
$. \end{itemize}
\par
The probability density $P(N_1,\dots, N_i, N_j,\dots,t)$ is given by the so-called master 
equation
\begin{equation}\label{eqf9}
\frac{\partial P}{\partial t} = W P
\end{equation}
which can be solved analytically only in some  very special cases \cite{vankampen}.
\par 
 
{\bf (2) The master equation as a model of scientific productivity}\\
\begin{figure}[ht]
\begin{center}
\includegraphics[scale=0.34]{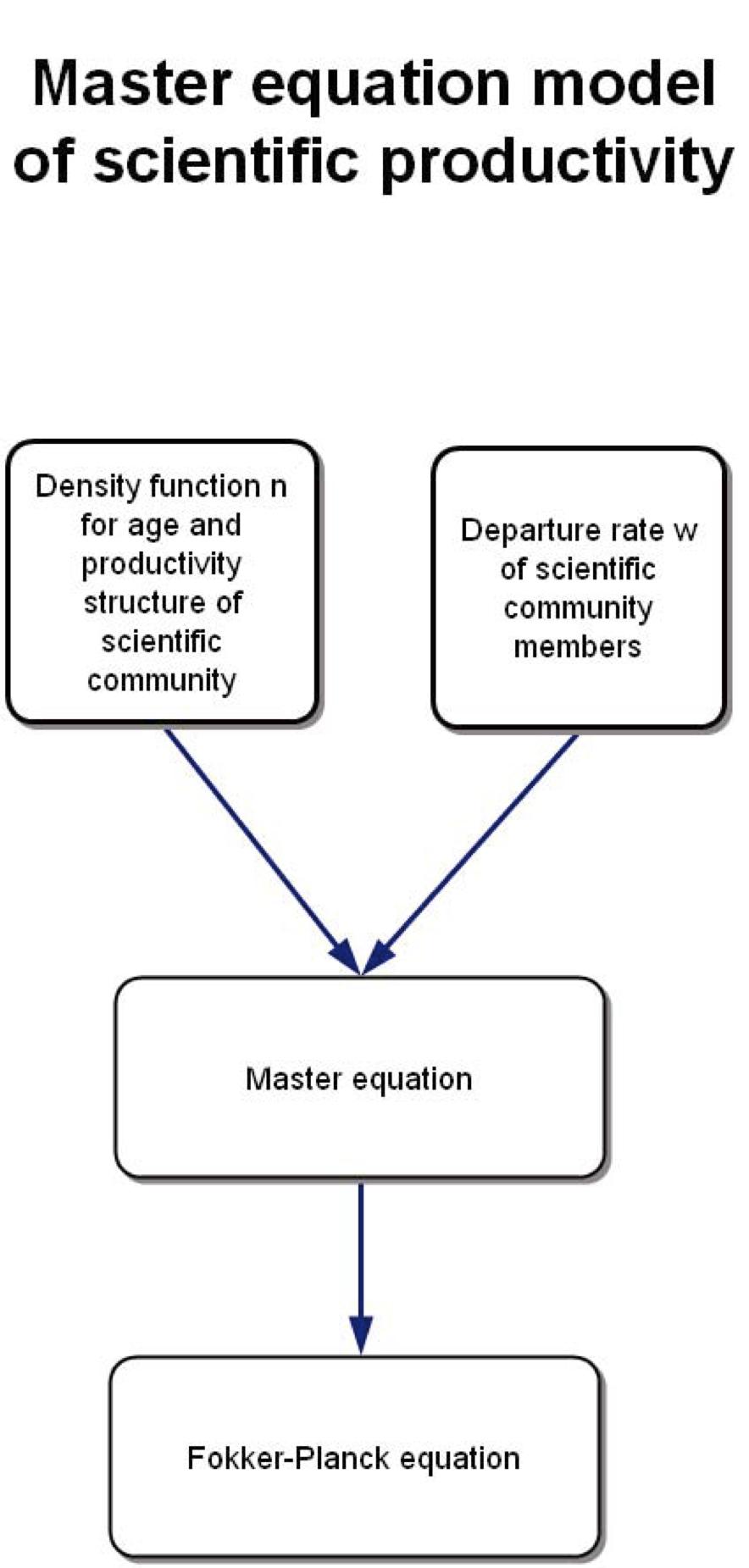}
\end{center}
\caption{Schema of the master equation model for scientific productivity.}
\end{figure}
The productivity factor is  a very important ingredient in mathematically simulating a
 scientific community evolution. One way to model such an evolution is through  a dynamic equation
which takes into account the stochastic fluctuations of scientific community members 
productivity \cite{rom} (Fig. 20). The main processes of scientific community
evolution accounted for by this model are, beside the biological  constraints (like the 
self-reproduction, aging of scientists,  and death), their departure  from the field due to  
mobility or  abandon of  research activities.  Call  $a$  the age of an individual and let a 
scientific productivity index $\xi$  be in incorporated into the  individual state space; 
both $a$  and $\xi$ are being considered  to be continuous variables with values in  $[0,\infty]$. The 
scientific community dynamics  is described by a number
density function $n(a, \xi, t)$, - another form of scientific landscape,  which  specifies the 
age and productivity structure of the  
scientific community at time $t$.   For example, the number of individuals with age in 
$[a_1,a_2]$ and scientific productivity in $[\xi_1, \xi_2]$ at time $t$ is given 
by the integral $\int_{a_1}^{a_2} \int_{\xi_1}^{\xi_2} da\ d \xi \  n(a, \xi, t)$. 
\par
A  master equation for this function $n( a,\xi, t )$ can be derived \cite{rom}:
\begin{eqnarray}\label{eqe6}
\left(\frac{\partial}{\partial a} + \frac{\partial}{\partial t} \right) \ n( a,\xi, t ) =
-[J(a,\xi,t) + w (a, \xi, t)] \ n(a,\xi,t) + \nonumber \\
\int_{-\infty}^{\xi} d \xi'  \ \chi(a, \xi - \xi',t) \ n(a,\xi - \xi',t),
\end{eqnarray}
where
$w(a, \xi, t)$ denotes the departure rate of community members. If $x(t)$ is a
random process describing the scientific productivity variation and  if  $p_a(x, t \mid y, \tau)$ (with $\tau < t$) is
the transition probability density corresponding to such a process, then
\begin{equation}\label{eqe3}
\chi(a, \xi, \xi', t) = \lim_{\Delta t \to 0}\frac{p_a(\xi +\xi', 
t + \Delta t \mid \xi, t)}{\Delta t}.
\end{equation}
The transition rate, at time
$t$ from the productivity level $\xi$, $J(a, \xi, t)$ is by definition:
$J ( a,\xi,t )= \int_{-\xi}^{\infty} d \xi' \ \chi (a, \xi, \xi',t)$. The increment $\xi'$ 
may be  positive or negative. The balance equation for 
$n(a, \xi, t)$  reads as follows
\begin{eqnarray}\label{eqe5}
n(a + \Delta a,\xi,t + \Delta t) = n(a,\xi,t ) - J (a,\xi, t )\ n( a,\xi, t )\  \Delta t 
+  \nonumber \\
\left[ \int_{-\infty}^{\xi}\ \chi(a, \xi - \xi', t) \ n( a,\xi - \xi',t ) d\xi'  \right]  \ \Delta t - 
w( a,\xi,t )\ n( a,\xi, t )\  \Delta t .
\end{eqnarray}
\par
The term on the right-hand side, $[1-J(a, \xi, t) \Delta t] n(a, \xi, t)$, describes the  
proportion of individuals whose scientific productivity does not  change in [$t, t + \Delta 
t$]; the integral term describes the individuals whose scientific productivity becomes equal 
to $\xi$ because of increasing or decreasing in [$t, t + \Delta 
t$];   the last   term 
corresponds to the departure of individuals due to stopping  research activities or death. 
After expanding $n(a + \Delta t, \xi, t + \Delta t)$ around $a$ and $t$, keeping terms  
up to the first order in $\Delta t$, one obtains the master equation Eq.(\ref{eqe6}).
\par
As the master equation is  difficult  to handle for an elaborate analysis, it is often reduced to 
an approximated equation similar to the well-known Fokker-Planck equation \cite
{risken,ref15,gardiner}. The approximation goes as follows. Let 
\begin{equation}\label{eqe48}
 \mu_k (a,\xi, t) = \int_{-\xi}^{ \infty} d \xi' (\xi')^k \chi(a, \xi, \xi',t)=
\lim_{\Delta t \to 0}\frac{1}{\Delta t} <(\xi')^k> ; \hskip.15cm  k=1,2,\dots ,
\end{equation}
where the brackets denote the average with respect to the
conditional probability density $p_a(\xi+\xi', t + \Delta t \mid  \xi, t)$. In addition,
the following assumptions are made: (i) $\mu_1, \mu_2  < \infty$; $\mu_k = 0$ for $k > 3$;
 (ii) $n(a, \xi, t)$ and $\chi(a, \xi, \xi', t)$ are analytic in $\xi$ for all $a$, $t$ 
and $\xi'$. The additional assumption $\mu_k =0$ for $k>3$ demands the productivity to be 
continuous in  the sense that as
$\Delta  t \to 0$, the probability of large fluctuations $\mid \xi' \mid$ must decrease 
so quickly that $<\mid \xi' \mid^3> \to 0$ more quickly than $\Delta t$.
\par
When the above assumptions hold, the function $n$ satisfies the equation \cite{rom}:
\begin{equation}\label{eqe8}
\left(\frac{\partial}{\partial a} + \frac{\partial}{\partial t} \right) n =
-\frac{\partial(\mu_1 n)}{\partial \xi} + \frac{1}{2} \frac{\partial^2 (\mu_2 n)}{
\partial \xi^2}- wn.
\end{equation}
If $w=0$, Eq.(\ref{eqe8}) is converted to the well known Fokker-Planck equation. 
Eq.(\ref{eqe8}) describes the
scientific community evolution through a drift along the age component and a drift
and diffusion with respect to the  productivity component. The diffusion term
characterized by the diffusivity $\mu_2$ takes into account the stochastic fluctuations of
scientific productivity conditioned by internal factors (such as individual abilities,
labour motivations, etc.) and external factors (such as labor organization, stimulation
system, etc.). The initial and boundary conditions for Eq.(\ref{eqe8}) are: (a)
$n(a,\xi,0) = n^0 (a, \xi)$,
where $n^0(a,\xi)$ is a known function defining the community age and productivity
distribution at time $t=0$; and (b) $n(0,\xi, t) =  \nu (\xi,t)$ 
where the function $\nu(\xi, t)$ represents the intensity of input flow of new 
members at age $a=0$ being set $\nu (\xi, 0) = n^0(0,\xi)$. In addition,
$n(a, \xi, t) \to 0$ as $a \to \infty$. 
\par
The general solution of equation Eq.(\ref{eqe8}) with the above initial condition (a) 
and boundary condition (b) is  still a difficult task. 
However, for many practical applications, a knowledge of first and second moments of 
distribution function $n(a, \xi, t)$ is sufficient. Eq.(\ref{eqe8}) can be solved 
numerically or can be reduced to a system
of ordinary differential equations \cite{rom}.
\begin{center}
{\fbox{\fbox{\parbox{13.0cm}{ \vskip0.5cm \sf 
FOR POLICY-MAKERS\\
 Take away box Nr.12: \\In   deterministic cases, the system  is robust against fluctuations:
it follows some trajectory and the fluctuations are too weak to change it. When the 
fluctuations
are important, then different trajectories for the evolution of the system become possible.
To each trajectory, a probability can be assigned. This probability reflects
the chance that the system will follow the corresponding trajectory. The collection of the
probabilities leads to a probability distribution which can be 
calculated,  in many evolutionary cases, on  the basis of the master equation approach.   
 \vskip0.5cm }}}}\\
\end{center}
\par
Finally,  two additional problems  that can be treated by the master equation approach can be mentioned: 
\begin{itemize}
\item
Age-dependent models where the birth and death rates connected to the
selection are age-dependent \cite{ex0, ex01}
\item
The problem of new species in evolving networks \cite{ax22}. On the
basis of a stochastic treatment of the problem,  the  notion of  'innovation'  can be
introduced in a broad sense as a disturbance and/or an  instability of  a corresponding  social, technological,  or scientific system.
The fate of a small number of individuals of a new species in a biological system
can be thought to be mathematically equivalent to some extent to the fate of  a new idea, a new
 technology, or a new model of behavior. The evolution of the new species can be studied on evolving networks, where some nodes can disappear and new nodes can be introduced. This evolution of the network can change significantly
the dynamic behavior of the entire system of interacting species itself. Some of the
species can vanish in a finite time. This feature can be captured effectively by the
master equation approach. 
\end{itemize}
\section{Space-time models. Competition of ideas. Ideological struggle}
\begin{figure}[h]
\begin{center}
\includegraphics[scale=0.44]{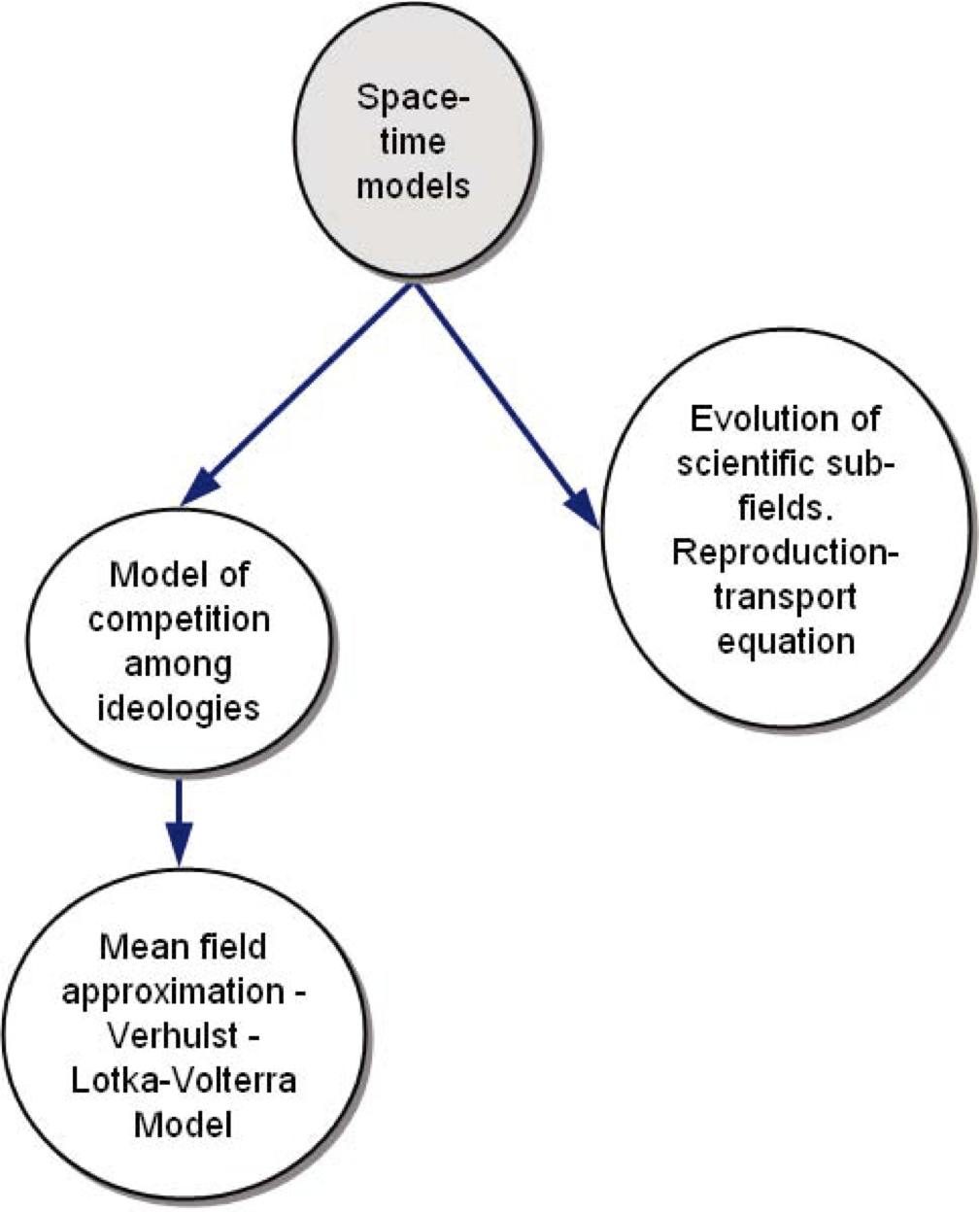}
\end{center}
\caption{Relations between space-time models discussed in this chapter.}
\end{figure}
A further level of complication is to include spatial variables explicitly in the  
above models  describing  the diffusion of ideas. At this stage of globalization of 
economies, with  several of its concomitant  features, like idea, knowledge,  and technology  diffusion,  to consider the spatial aspect is clearly 
a must.  A large amount of research on the
spatial aspects of diffusion of populations is already  available. As examples of early work,  papers by Kerner \cite{kerner1},  Allen \cite{allen75}, Okubo \cite{okubo80},
and Willson and de Roos \cite{willson93} can be pointed out. 
From the point of view of diffusion of ideas and
scientists,  the previously discussed continuous model of research mobility  \cite{andrea1} 
has to 
be singled out. Moreover, the model presented below is closely connected to the space-time 
models of migration of populations developed by Vitanov and co-authors \cite{dv091,dv092}. In 
addition, a reproduction-transport equation model
(see Fig. 21) can be discussed.
\subsection{Model of competition between ideologies }
\begin{figure}[h]
\begin{center}
\includegraphics[scale=0.4]{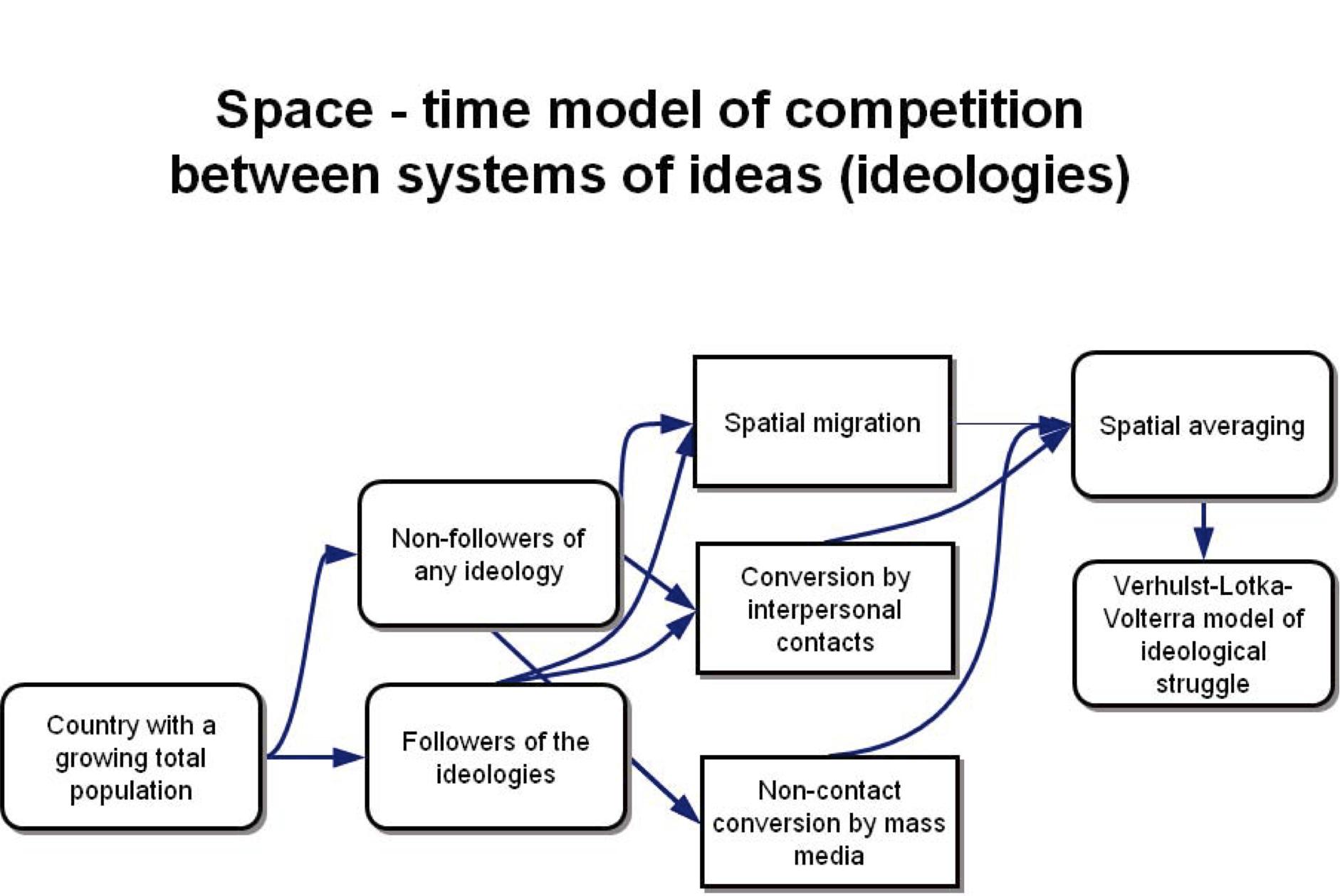}
\end{center}
\caption{Schema of the space-time model for describing competition between ideologies.}
\end{figure}
The diffusion of ideas is necessarily  accompanied by competition processes.
One model of competition between systems of ideas (ideologies)  goes as follows (Fig. 22).
Let a population of $N$ individuals occupy a two-dimensional plane.
Suppose that there exists a set of ideas or ideologies $P=\{P_{0},P_{1}, \dots, P_{n} \}$
and  let  $N_i$ members of the population be  followers of the $P_i$  ideology .  
The members $N_0$ of the class  $P_{0}$ are not supporters of any ideology; in some 
sense, they have their own individual one and do not wish to be considered associated 
with another one, global or not. 
In such a way, the population is divided in $n+1$ sub-populations
of followers of different ideologies. The total population is: 
$N = N_{0} + N_{1} + \dots N_{n}$.
Let  a small region $\Delta S = \Delta x \Delta y$ be selected  in the plane.
In this region  there are $\Delta N_{i}$ individuals  holding the $i$-th
ideology, $i=0,1,\dots,n$. If  $\Delta S$ is sufficiently small, the density of 
the $i$-th population can be defined as 
$\rho_{i} (x, y, t) = \frac{\Delta N_{i}}{\Delta S}$.
\par
Allow the members of the $i$-th population to move through the borders
of the area $\Delta S$. Let $\vec{j}_{i}(x,y,t)$ be the
current of this movement. Then $(\vec{j}_{i} \cdot \vec{n}) \delta l$
is the net number of  members of the $i$-th population/ideology, crossing
a small line $\delta l$ with normal vector $\vec{n}$.  
Let the changes be summarized by the function 
$C_{i}(x,y,t)$. The total change of the number of members of the $i$-th population  is
\begin{equation}\label{diff_eq}
\frac{\partial \rho_{i}}{\partial t} + {\rm div} \vec{j}_{i} = C_{i}.
\end{equation}
The first term in Eq.(\ref{diff_eq}) describes the net rate of increase of the
density of the $i$-th population. The second term describes the net rate of
immigration into the area. The  r.h.s. of Eq.(\ref{diff_eq}) describes the 
net rate of increase exclusive of immigration.  
\par
Let us now specify $\vec{j}_{i}$ and $C_{i}$:  $\vec{j}_{i}$ is assumed
to be made of a non-diffusion part $\vec{j}_{i}^{(1)}$ and a diffusion 
part $\vec{j}_{i}^{(2)}$ where  $\vec{j}_{i}^{(2)}$ is 
assumed to have the general form of a linear multicomponent diffusion 
\cite{kerner1} in terms of a diffusion coefficient $D_{ik}$
\begin{equation}\label{current}
\vec{j}_{i} = \vec{j}_{I}^{(1)} + \vec{j}_{2}^{(2)} =
\vec{j}_{i}^{(1)} - \sum_{k=0}^{n} D_{ik} (\rho_{i}, \rho_{k},x, y, t) \nabla \rho_{k}.
\end{equation}
Let some of the followers of the ideology $P_{i}$  be capable of and interested 
in changing  ideology:  i.e., 
they can convert from the ideology $P_{i}$ to the ideology $P_{j}$. 
It can be  assumed that the following processes can happen with respect to 
the members of the subpopulations of the property holders:
{\bf (a)} deaths: described by a term $r_{i} \rho_{i}$.  It is assumed  that the  
number of deaths in the $i$-th population is proportional to its population density. 
In general  $r_{i}=r_{i}(\rho_{\nu}, x, y, t; p_{\mu})$,  where  $\rho_{\nu}$ stands  
for ($\rho_{0}, \rho_{1}, \dots, \rho_{N}$)
 and  $p_{\mu}$ stands  for $(p_{1},\dots, p_{M})$   containing  parameters of 
the environment; 
{\bf (b)} non-contact  conversion: in this class  are  included all conversions
exclusive of the conversions by interpersonal contact between the members
of  whatever populations. A reason for non-contact conversion can be the existence 
of different kinds  of
 mass communication  media which make propaganda  for  whatever  ideologies.
As a result,  members of each population can change ideology. For
the $i$-th population, the  change in  the number of members is: $\sum_{j=0}^{n} f_{ij} 
\rho_{j}$, $f_{ii}=0$. In general,
$f_{ij}=f_{ij}(\rho_{\nu}, x, y, t; p_{\mu})$;
{\bf (c)} contact conversion: it is   assumed that there can be interpersonal contacts 
among the population members.
The contacts happen  between members in groups consisting of two members (binary contacts),
three members (ternary contacts), four members, etc. 
As a result of the contacts,  members of  each population can change
their ideology. For binary contacts, let it be assumed that the  change of
ideology  probability for a member of the $j$-th population is proportional to the
possible number of contacts,  i.e.,  to the density of the $i$-th population.
Then the total number of "conversions" from $P_{j}$ to $P_{i}$
is $a_{ij} \rho_{i} \rho_{j}$, where $a_{ij}$ is a parameter. 
In order to have a ternary contact, one 
must have a group of three members. The most simple is to assume that  such a group exists 
with a probability proportional to the corresponding densities of the concerned populations. 
In a  ternary contact between  members of the $i$-th, $j$-th, and $k$-th population, 
  members of the $j$-th and $k$-th populations can change their ideology according to
$P_{i}$ = $b_{ijk} \rho_{i} \rho_{j} \rho_{k}$, where $b_{ijk}$
is a parameter. In general,
$
a_{ij}=a_{ij}(\rho_{\nu}, x, y, t; p_{\mu})
$;
$
b_{ijk} = b_{ijk}(\rho_{\nu}, x, y, t; p_{\mu})
$; etc.
\par
On the basis of the above, the  $C_{i}$  term looks as follows
(for more research of these types of population models see \cite{dimvit00, dimvit01, dimvit01a}):
\begin{equation}
 C_{i}=r_{i} \rho_{i}  + \sum_{j=0}^{n} f_{ij} \rho_{j}
+ \sum_{j=0}^{n} a_{ij} \rho_{i} \rho_{j} + \sum_{j,k=0}^{n}
b_{ijk} \rho_{i} \rho_{j} \rho_{k}+ \dots ,\end{equation}
 and the model system of equations becomes
\begin{eqnarray}\label{model01}
\frac{\partial \rho_{i}}{\partial t} + {\rm div} \vec{j}_{i}^{(1)} - 
\sum_{j=0}^{n} {\rm div} (D_{ij} \nabla \rho_{j}) =
r_{i} \rho_{i}  + \sum_{j=0}^{n} f_{ij} \rho_{j}
+ \nonumber \\
\sum_{j=0}^{n} a_{ij} \rho_{i} \rho_{j} + \sum_{j,k=0}^{n}
b_{ijk} \rho_{i} \rho_{j} \rho_{k}+ \dots
\end{eqnarray}
The density of the entire population is
$\rho = \sum_{i=0}^{n} \rho_{i}$.
It can be  assumed that it  changes in time according to the Verhulst law (but see 
the note after Eq.(\ref{modelx3})!)
\begin{equation}\label{verhulst}
\frac{\partial \rho}{\partial t} = r \rho \left( 1- \frac{\rho}{C} \right)
\end{equation}
where $C(\rho_{\nu}, x, y, t; p_{\mu})$ 
is the so-called carrying capacity of the environment  \cite{Odum59}
and $r(\rho_{\nu}, x, y, t; p_{\mu})$
is a positive or negative growth rate. When pertinent sociological data are available, the 
same type of equation could hold for any $i$-th population with a given $r_i$.
\par
First,  consider the case in which the current $\vec{j}_{i}^{(i)}$ is negligible, i.e.,
$\vec{j}_{i}^{(i)} \approx 0$ ($no$ diffusion approximation).
In addition,  consider only the case when all parameters are  constants. 
The model system of equations becomes 
\begin{eqnarray}\label{modelmarc}
\frac{\partial \rho_{i}}{\partial t}  - D_{ij} 
\sum_{j=0}^{n}  \Delta \rho_{j} =
r_{i} \rho_{i} + \sum_{j=0}^{n} f_{ij} \rho_{j}
+ \sum_{j=0}^{n} a_{ij} \rho_{i} \rho_{j} + \nonumber \\
 \sum_{j,k=0}^{n} b_{ijk} \rho_{i} \rho_{j} \rho_{k}+ \dots,
\hskip.25 cm  \end{eqnarray} \\   for \begin{eqnarray} \;\;\;\;  \Delta = \frac{\partial^{2}}{
\partial x^{2}} + 
\frac{\partial^{2}}{\partial y^{2}}, \;\;\; \hskip.35cm i=0,1,2,\dots,n.
\end{eqnarray}
\par
Let  plane-averaged quantities and fluctuations
(linear or nonlinear) be enough relevant. Let $q(x,y,t)$ be a quantity defined in
an area $S$. By definition, a plane-averaged quantity is
$\overline{q} = \frac{1}{S} \int \int_{S} dx dy \ q(x,y,t)$.
Call the  fluctuations $Q(x,y,t)$  such that  $q(x,y,t)= \overline{q}(t) + Q (x,y,t)$.
If  the  territory is large and within the stationary 
approximation, $S$  can be assumed to be large enough such that each
plane-averaged combination of fluctuations vanishes,  such that 
$\overline{Q_{i}}= \overline{Q_{i} Q_{j}} = \overline{Q_{i} Q_{j} Q_{k}} =
\dots = 0$.
In addition to $S$  being large and $\int \int_{S} dx dy \Delta Q_{k}$ 
 assumed to be  finite,  it can be also assumed  that 
$\overline{\Delta Q_{k}} = \frac{1}{S} \int \int_{S} dx dy \Delta Q_{k}
\to 0$.
\par
On the basis of the above (reasonable) assumptions, it is possible to separate the dynamics
of the averaged quantities from the dynamics of fluctuations.   As a result of  the  plane-average of Eq.(\ref{modelmarc}), the following equations for the dynamics of the plane-averaged densities are obtained 
\begin{equation}\label{modelx1}
\overline{\rho}_{0} = \overline{\rho} - \sum_{i=1}^{n} \overline{\rho}_{i}; \hskip.3cm
\frac{d \overline{\rho}}{d t} = r \overline{\rho} \left(1 - 
\frac{\overline{\rho}}{C} \right)
\end{equation}
\begin{equation}\label{modelx3}
\frac{d \overline{\rho}_{i}}{dt} = r_{i} \overline{\rho}_{i} +
\sum_{j=0}^{n} f_{ij} \overline{\rho}_{j} + \sum_{j=0}^{n} a_{ij} 
\overline{\rho}_{i} \overline{\rho}_{j} + \sum_{j,k=0}^{n} b_{ijk}
\overline{\rho}_{i} \overline{\rho}_{j}\overline{\rho}_{k} + \dots
\end{equation}
Instead of (\ref{modelx1}) we can write an equation for $\overline{\rho}_0$
from the kind of (\ref{modelx3}).  Then the total population density $\overline{\rho}$
will not follow the Verhulst law. 
\par
Equations (\ref{modelx1}) and (\ref{modelx3})  represent the model of ideological
struggle proposed by Vitanov, Dimitrova and Ausloos \cite{vda}. There is one important
difference between the Lotka-Volterra models  \cite{lotka,volterra}, often used for
describing prey-predator systems, and the above model of ideological struggle.  
The originality resides in the generalization of usual prey-predator models to the case  in 
which a prey (or predator) changes its state and becomes a member of the predator pack 
(or prey band), due to some interaction with its environment or with some
other prey or predator. Indeed, it can be hard for rabbits and foxes to do so, but it can be 
often the case in a society: a member of one population can drop his/her ideology and can convert 
to another one.
\par
In order to show the relevance of such extra conditions on an evolution of populations, 
consider a huge (mathematical) approximation, - it might be a drastic one in particular in  a  country with  a  strictly 
growing total population.  (Recall that the  growth rate $r$ could be positive or negative or 
time-dependent).  Let $r $ be  $>0$ and let 
the maximum possible population of the country  be $C$.
Consider  more convenient notations by setting $\overline{\rho} = N$;
$\overline{\rho}_0 = N_0$; $\overline{\rho}_i = N_i$ and assume that the binary contact 
conversion is much stronger than the ternary, etc. conversions. The system  
equations become
\begin{equation} \label{model1}
N = N_{0} + \sum_{i=1}^{n} N_{i}; \hskip.5cm
\frac{dN}{dt}=r N  \left( 1 - \frac{N}{C} \right) 
\end{equation}
\begin{eqnarray}\label{model3}
\frac{dN_{i}}{dt} = 
r_{i}N_{i}+ 
\sum_{j=0}^{n} f_{ij} N_{j} + 
\sum_{j=0}^{n} b_{ij} N_{i} N_{j}.
\end{eqnarray}
\par
Reduce the discussion  of Eqs.(\ref{model1}) and (\ref{model3}) to a  society  in which there 
is  the  spreading of only one ideology; therefore,  
the population of the country  is  divided into two groups: $N_{1}$, followers of 
the "invading"  ideology   and $N_{0}$, people who are at first "indifferent" to this ideology.
Let  only the non-contact conversion scheme exist,  as possibly 
moving the  ideology-free population toward the single ideology; thus $f_{10}$ is finite,
but $b_{10}=0$. Let the initial conditions be  $N(t=0)=N(0)$
and $N_{1}(t=0)=N_{1}(0)$. 
The solution of the  system of model equations is
\begin{equation}\label{solution1}
N(t) = \frac{C N(0)}{N(0) + (C-N(0))e^{-rt}},
\end{equation}
like the Verhulst law, but
\begin{eqnarray}\label{solution2}
N_{1}(t) = e^{-(f_{10}-r_{1})t} \bigg \{ N_{1}(0) + \frac{C f_{10}}{r} 
\bigg [ 
{\Phi}\bigg ( -\frac{C-N(0)}{N(0)},1,-\frac{f_{10}-r_{1}}{r} \bigg ) -
\nonumber \\
e^{(f_{10}-r_{1})t} {\Phi} \bigg(-\frac{C-N(0)}{N(0) e^{rt}},1,
- \frac{f_{10}-r_{1}}{r}\bigg) \bigg] \bigg\}
\end{eqnarray}
with 
\begin{equation}\label{solution3}
N_{0}(t)=N(t)-N_{1}(t)
\end{equation}
in which $\Phi$ is the special function $
{\Phi} (z,a,v) = \sum_{n=0}^{\infty} \frac{z^{n}}{(v+n)^{a}} \hskip.1cm ; \hskip.5cm 
\mid z \mid < 1$.
\par
The obtained solution describes an evolution in which the total population $N$ reaches
asymptotically the carrying capacity $C$ of the environment. The number of adepts of
the ideology reaches an equilibrium value which
corresponds to the fixed point $ \hat{N}_{1} = C f_{10}/(f_{10} -r_{1})$ of 
the model equation for $\frac{d N_{1}}{dt}$.
The number of people who are not followers of the ideology asymptotically
tends to  $N_{0}=C-\hat{N}_{1}$.
Let  $C=1$, $f_{10}=0.03$, and $r_{1}=-0.02$, then $\hat{N}_{1}=0.6$, which
means that the evolution of the system leads to an asymptotic state in which
60 \% of the population are followers of the ideology and 40 \%  are not. 
\par
Other more complex cases  with several competing ideologies can be discussed, observing 
steady states or/and cycles (with different values of the time intervals for each growth or/and decay), 
chaotic behaviors, etc. \cite{vda}. In particular,  it  can be shown that  accepting a slight 
change in the conditions of the environment can prevent the
extinction  of some ideology. After almost collapsing, some 
ideology can spread again and  can affect a significant part of the country's population. Two
kinds of such resurrection effects have been found and described  as {\it phoenix effects} in 
the case of two competing ideologies. In the phoenix effect of the first kind, the 
equilibrium state connected to the extinction of the second
ideology exists but is unstable. In the phoenix effect of  the so-called second kind, the 
equilibrium state connected to extinction of the second ideology vanishes.  {\it In fine}, 
the above model seems powerful enough to discuss many realistic cases. The number of control 
parameters seems huge, but that is the case for many   competing epidemics in complex systems. 
However, it was observed that the values of parameters can be monitored when enough data is 
available, including the time scales \cite{vda}.
\begin{center}
{\fbox{\fbox{\parbox{13.0cm}{ \vskip0.5cm \sf 
FOR POLICY-MAKERS\\
 Take away box Nr.13: \\Space-time models are very appropriate for modeling
migration processes such as the spatial migration of scientists, besides the diffusion of 
ideas through competition  without strictly physical motion.   
 \vskip0.5cm }}}}\\
\end{center}
\subsection{Continuous model of evolution of scientific subfields. Reproduction-transport 
equation}
\begin{figure}[h]
\begin{center}
\includegraphics[scale=0.35]{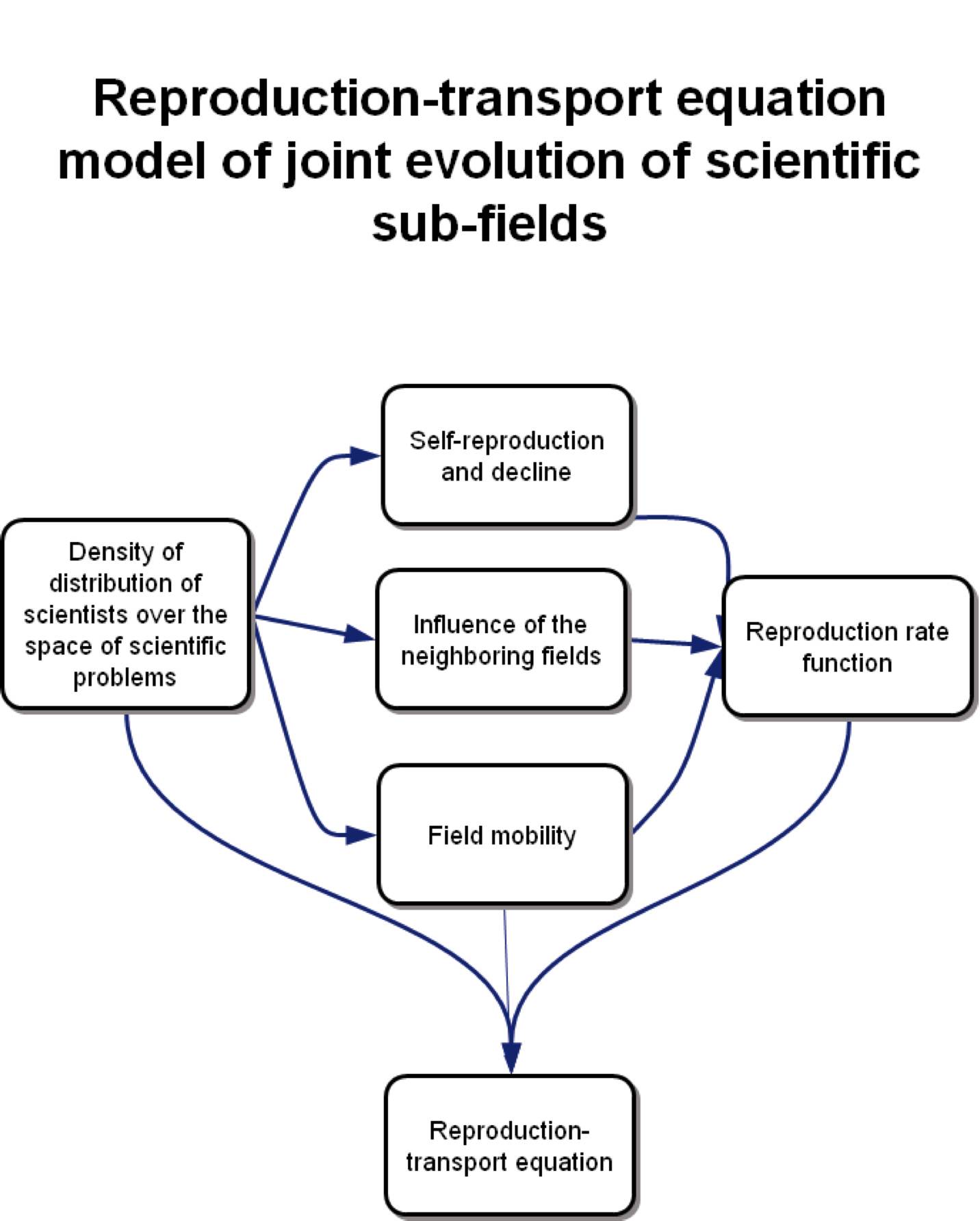}
\end{center}
\caption{Schema of the reproduction-transport equation model of joint evolution of
scientific fields.}
\end{figure}
The change of subject of a scientist can be  considered as a migration process 
\cite{andrea1,ebeling00}.  Let research problems be represented
by sequences of signal words or macro-terms
$P_i = (m_i^1, m_i^2,\dots, m_i^k, \dots, m_i^n)$
which are registered according to the frequency of their appearance, joint appearance, etc., respectively, 
 in the texts. Each point of the problem space, described by a vector 
$\vec{q}$, corresponds to a research problem, with the problem space consisting of all 
scientific problems (no matter whether they are under investigation or not). The scientists 
distribute themselves over the space of scientific problems with density $x(\vec{q},t)$. Thus, there is a number $x(\vec{q},t)d\vec{q}$ working at  time $t$ in the element $d \vec{q}$.
The field mobility processes correspond to a density change of scientists in the problem
space: instead of  working on problem $\vec{q}$, a scientist may  begin  to work on 
problem $\vec{q'}$. As a result, $x(\vec{q},t)$ decreases and $x(\vec{q'},t)$  increases. This 
movement of scientists (see also Fig. 23) can be described by means of the following  
reproduction-transport-equation:
\begin{equation}\label{eqf11}
\frac{\partial x(\vec{q},t)}{\partial t} = x (\vec{q},t) \ w(\vec{q} \mid x) + 
\frac{\partial}{\partial 
\vec{q}} \left( f(\vec{q},x) + D(\vec{q}) \frac{\partial x (\vec{q},t)}{\partial \vec{q}} \right).
\end{equation}
In  Eq.(\ref{eqf11}), self-reproduction and 
decline are represented by the term $w(\vec{q} \mid x) \  x(\vec{q}, t)$. For the 
reproduction rate  function $w(\vec{q} \mid x)$, one can write
\begin{equation}\label{eqf12}
w(\vec{q} \mid x) = a(\vec{q}) + \int d \vec{q'} \ b(\vec{q}, \vec{q'}) \  x(\vec{q'},t).
\end{equation} 
The local value of $a(\vec{q})$ is an expression of the rate at which the number of 
scientists on field $\vec{q}$ is modified through self-reproduction and/or decline, while 
$b(\vec{q},\vec{q'})$ describes
the influence exerted on the field $\vec{q}$ by the neighbouring field $\vec{q'}$.
The field mobility is modeled by means of the term $\frac{\partial}{\partial \vec{q}}
\left(f(\vec{q},x) + D(\vec{q}) \frac{\partial}{\partial \vec{q}} x(\vec{q},t) \right).$
In  most cases, Eq.(\ref{eqf11}) can only  be solved numerically. For more details
on the model, see \cite{andrea1}. 
\section{Statistical approaches to the diffusion of knowledge}
\begin{figure}[h]
\begin{center}
\includegraphics[scale=0.55]{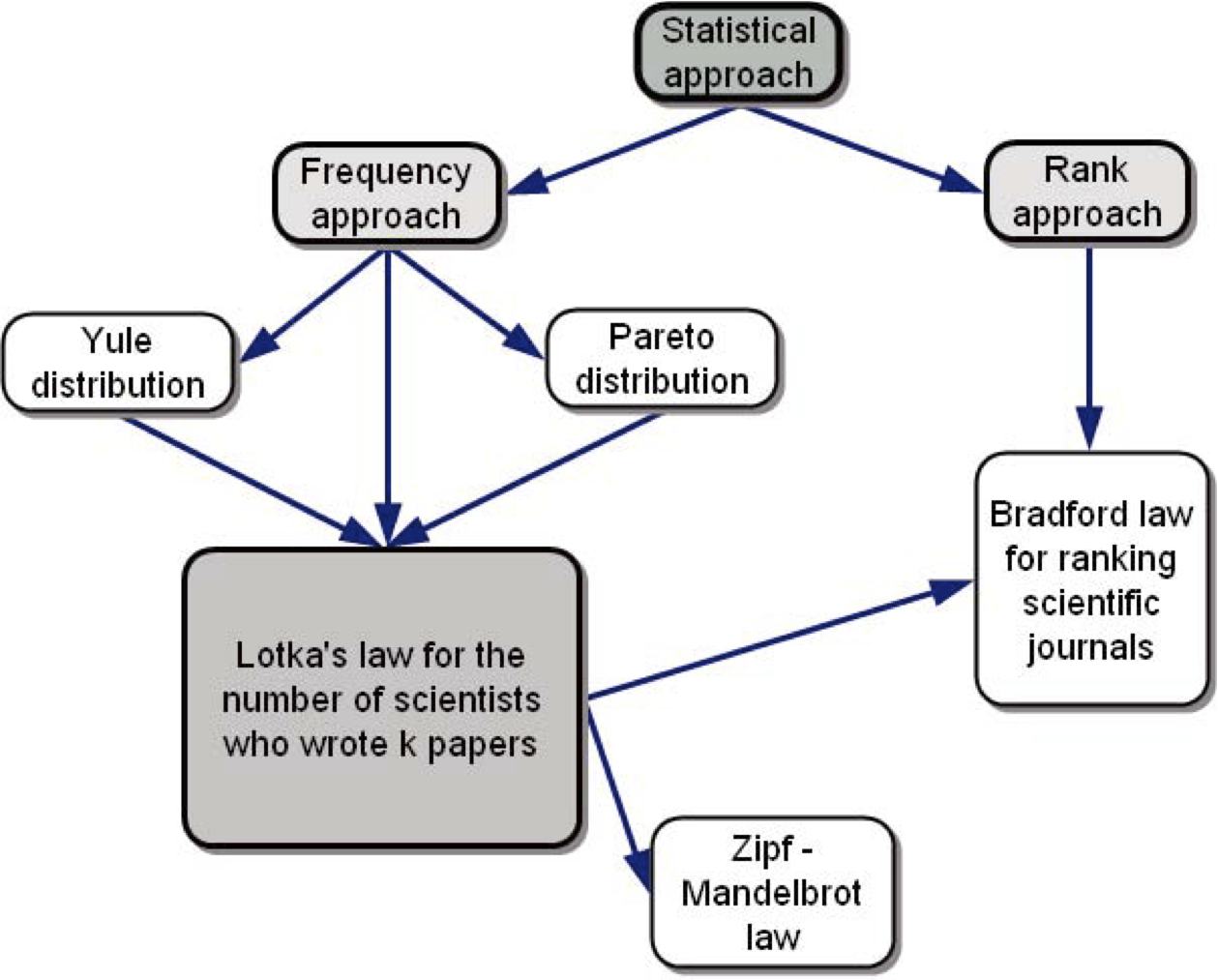}
\caption{Statistical laws and their relationships as discussed in the chapter.}
\end{center}
\end{figure}
 Solomon and Richmond \cite{sr,sr1} have shown that
the systems of generalized Lotka-Volterra equations   are closely connected to the
Pareto-Zipf  probability  distribution.  Since  such a distribution arises among 
other distributions and laws connected to the description of the diffusion of knowledge,  it 
is of interest to discuss briefly the diffusion of knowledge within statistical approach studies. 
Lotka was its  pioneer; a large amount of research   has followed. Just as examples,  one can mention the work of Yablonsky and Haitun 
 on the Lotka  law for the distribution of 
scientific productivity and its connection with the Yule distribution \cite
{yab1,yab2,haitun82}, where the non-Gaussian nature of the scientific activities is 
emphasized.  Interesting applications of the Zipf  law are also presented in \cite{li02}. 
The connection  to the non-Gaussian distributions concepts of self-similarity and 
fractality have been applied to the scientific system in \cite{katz99} and \cite{raan00}. 
 Several tools for appropriate statistical analysis  are hereby discussed. At the center
of the discussion   Lotka  law shall receive some special attention (see Fig. 24) \footnote{
Let us mention a curious and interesting fact connected to statistical indicators.
Very interesting is the conclusion in \cite{gao09} that the scale-independent
indicators show that  in the fast growing innovation system of China, research institutions
financed by the government play a more important role than the enterprises.}.   
\par
As part of this discussion on the statistical approach,  the analysis of the
productivity of scientists can be considered. The information connected to new ideas is
 thought to be often  codified in scientific papers. Thus,  the statistical aspects of 
scientific productivity is of practical importance.  For  example, the Lotka law reflects the 
distribution of publications over the set of   authors considered as the information
sources.   Bradford  law describes the distribution of papers on
a given topic over the set of  journals publishing these
papers and ranked  according to  the order in  the decrease of the number of papers on a 
given topic in each journal.  These laws have a non-Gaussian nature and, because of
this,  possess specific features such as a concentration
and dispersal effect \cite{yab1}:  for  example, it is found that there is a small number
of highly productive scientists who write most of the papers on a given topic and, on the 
other hand, a large number of scientists with low productivity.
\par
In order to give an example of the connection between the deterministic and statistical
approaches, remember that the   Goffman-Newill model, discussed here above,  presents 
a connection between  the number of scientists working in a research area and the 
number of relevant publications. In \cite{bett2}, it was found that the number of new
publications scale as a simple power law with the corresponding number of new
authors: $\Delta P  = C(\Delta T)^\alpha$
where $\Delta P$ and $\Delta T$ are the new publications and the new authors 
over some time period (for an example one year).  $C$ is a normalization constant, 
and $\alpha$ is  a  scaling exponent. It has been demonstrated \cite{bett2} that the latter  
relationship provides a very good fit to data for  six different research fields, but with 
different values of the scaling exponent $\alpha$. For $\alpha > 1$,  a field would
grow by showing an increase in the number of publications per capita, i.e., in such a research 
field,  the individual productivity increases as the field attracts new scientists. 
A field with $\alpha < 1$ has a per capita decrease in productivity. 
This can be a warning signal for a dying subject matter. It would be interesting to observe 
whether the exponent $\alpha$ is time-dependent, as is the case in  related characterizing 
scaling exponents of financial markets \cite{vdwMA279} or in meteorology\cite{KIMA349}. 
Policy control can thus be implemented for shaking $\alpha$, thus the field mobility.
\begin{center}
{\fbox{\fbox{\parbox{13.0cm}{ \vskip0.5cm \sf 
FOR POLICY-MAKERS\\
 Take away box Nr.14: \\There exist two different kinds of statistical approaches for  the 
analysis of scientific productivity:  (i)  the 
frequency approach and  (ii) the rank approach. The frequency approach is based on
the direct statistical counting of the number of   corresponding information
sources, such as scientists or journals. The rank approach is based on a ranking of 
the sources with respect to their productivity. The frequency and the  rank approaches  
represent different and complementary reflections of the same law and form.
 \vskip0.5cm }}}}\\
\end{center}
\subsection{Lotka  law. Distributions of Pareto and Yule.} 
Pareto \cite{chen93} formulated the 80/20 rule: it can be expected that 20\% 
of people will have 80\% of the wealth. Or it can be expected that 80\% of the citations 
refer to a core of 20\% of the titles in journals. The idea of the rule of Pareto is very 
close to the research of Lotka who noticed the  following dependence for the number of 
scientists $n_k$ who wrote $k$ papers  
\begin{equation}\label{lotka2}
n_k= \frac{n_1}{k^2}; \hskip.25cm k=1,2,\dots,k_{max}.
\end{equation}
In Eq.(\ref{lotka2}), $n_1$ is the number of scientists who wrote just one paper  and 
$k_{max}$ is the maximal productivity of a scientist.
\begin{equation}
 \sum_{k=1}^{k_{max}} n_k = n_1 \sum_{k=1}^{k_{max}}\frac{1}{k^2} =N
\end{equation}
where $N$ is  the total number of scientists. If we assume that $k_{max} \to \infty$  
and take into account the fact that $\sum_{k=1}^{\infty} 1/k^2 = \pi/6$,
we obtain a limiting value for the portion of scientists with the minimal productivity
(single paper authors) in the given population of authors: $P_1 = n_1/N \approx 0.6$.
Then, if the left and the right hand sides of Eq.(\ref{lotka2}) are
divided by N, the frequency expression for the productivity distribution is:
$p_1 = 0.6/k^2$; $\sum_{k=1}^{\infty} p_k =1$.
Eq. (\ref{lotka2}) is called Lotka law, or the law of inverse squares: the number of 
scientists who wrote a given number of papers is inversely
proportional to the square of this number of papers. 
\par
It must be noted that, like many other statistical regularities, Lotka 
law is valid only on the average since the exponent in the denominator of Eq.(\ref{lotka2})
is not necessarily equal to two \cite{yab1}. Thus, Lotka law should be 
considered as the most typical among a more general family of distributions:
\begin{equation}\label{lotka3}
n_k = \frac{n_1}{k^{1 + \alpha}}; \hskip.5cm p_1 = \frac{p_1}{k^{1 + \alpha}} 
\end{equation}
where $\alpha$ is the characteristic exponent of the distribution, 
$n_1$ is the normalizing coefficient which is determined as follows:
\begin{equation}\label{lotka4}
p_1 = \frac{n_1}{N} = \left (\sum_{k=1}^{k_{max}} \frac{1}{1+ k^{\alpha}} \right)^{-1}.
\end{equation}
Then the distribution of scientific output, Eq.(\ref{lotka3}), is determined 
by three parameters: the proportion of scientists with the minimal productivity
 $p_1$, the maximal productivity of a scientist $k_{max}$, and the
characteristic exponent $\alpha$. If one of these parameters is fixed, it is possible to
study the dependence between two others.  Let us fix 
$k_{max}$ in Eq.(\ref{lotka4}). Then, we obtain the proportion of "single paper authors" $p_1$
 as a function of $\alpha$: $p_1(\alpha)$. When Eq.(\ref{lotka4}) is differentiated
with respect to $\alpha$, one can show that the corresponding derivative is
positive for any $\alpha$ :
$dp_1(\alpha)/d \alpha > 0$. On the basis of a similar analysis of the portion of scientists with
a larger productivity $p_k(\alpha)$ as a function of $\alpha$, we arrive at the 
conclusion: {\bf the increase of $\alpha$ is accompanied by the increase of low-productivity scientists}. This means that when the total number of scientists is preserved
the portion of highly productive scientists will decrease. 
\par
 Let us show that the  Lotka law is  an asymptotic expression for the 
Yule distribution.    In order to obtain the Yule distribution, one  considers
the process of formation of a collection of publications as a
Markov-type stochastic process. In addition, it is  assumed 
that the probability of writing a new paper depends on the number of papers that
have been already written by the scientist at time $t$:  the
probability of the transition into a new state on the interval $[t, t + \Delta t]$ should be a
function of the state in which the system is at   time $t$.  Moreover, the probability
of publishing a new paper during a time  interval $\Delta t, p(x \to x + 1, \Delta t)$ is  
assumed to be proportional to the number $x$ of papers that have been written by the 
scientists, introducing an intensity coefficient $\lambda$: $p(x \to x+ 1, \Delta t) \propto 
\lambda  x \Delta t$.
After solving the corresponding system of differential equations for this process,  
the following expression (the Yule distribution) for
the probability $p(x/t)$ of a scientist writing $x$ papers during a time $t$ is obtained \cite
{yab1}:
\begin{equation}\label{yule1}
p(x/t) = \exp( -\lambda t) (1 - \exp(-\lambda t))^{x-1}, x = 1, 2 \dots
\end{equation}
The mean value of the Yule distribution is  $x_t = \exp({\lambda t})$.  
Let us take into account the fact that every scientist 
works on a given subject during a certain finite random time interval $[0,t]$  
which  depends on  the scientist's creative potential, the conditions for work, etc. 
With the simplest assumption that the probability of discontinuing
work on a given subject is constant at any  time, one obtains
an  exponential distribution for the time of work of any author in
the scientific field under study: $p(t) = \mu \exp( -\mu t)$, where $\mu$ is the 
distribution parameter. The  time parameter $t$ which characterizes the productivity
distribution, Eq.(\ref{yule1}), is a random number.   Then in order to obtain the
final distribution of scientific output observed in the experiment over sufficiently
large time intervals, Eq.(\ref{yule1}) should be averaged
with respect to this parameter $t$ which is distributed according to the exponential
law:
\begin{equation}\label{yule2}
p(x) = \int_0^{\infty} dt \  p(x/t) p(t)  = \int_0^{\infty} dt \  \exp(- \lambda t)
 (1-\exp(-\lambda t))\mu \exp(-\mu t).
\end{equation}
 After integrating Eq.(\ref{yule2}),   the distribution of scientific output reads
\begin{equation}\label{yule3}
p(x) = \frac{\mu}{\lambda}B \left(x, \frac{\mu}{\lambda} + 1 \right) = 
\alpha B (x, \alpha +1), x=1,2,\dots
\end{equation}
where $B(x, \alpha + 1) = \Gamma(x) \Gamma( \alpha x + 1)/ \Gamma(x + \alpha + 1)$ is 
a Beta-function, $\Gamma(x) \approx (x - 1) !$ is a Gamma-function, and 
$\alpha  = \mu/\lambda$ is the characteristic exponent.
For instance, if $\alpha \approx 1$ then $p(x) = 1/[x(x+1)]$.
Let us assume that $x \to  \infty$ and apply the Stirling formula. Thus,  the 
asymptotics of the Yule  distribution Eq.(\ref{yule3})  is like Lotka law Eq.(\ref{lotka3}) (up to 
a normalizing constant): 
$p(x) \propto \Gamma(\alpha + 1)\alpha/x^{1+ \alpha}$. 
\subsection{ Pareto distribution, Zipf-Mandelbrot and Bradford  laws}
For large enough values of the total number
of scientists and the total number of publications, we can make the transition from 
discrete to continuous  representation of the corresponding variables and laws.  
The continuous analog of Lotka law, Eq. (\ref{lotka3}), is the  Pareto  distribution
\begin{equation}\label{pareto1}
p(x) = \frac{\alpha}{x_0} \left(\frac{x_0}{x} \right)^{\alpha + 1}; \hskip.5cm
x \ge x_0; \hskip.25cm \alpha >0
\end{equation}
which  describes the distribution density
for a number of scientists with $x$ papers; $x_0$ is the minimal productivity
$x_0 << x << \infty$,  a continuous quantity. 
\par
 Zipf  law is connected to the principle of least effort \cite{z49}: 
a person will try to solve his problems in such a way as to minimize the total work 
that he must do in the solution  process. For  example, to express with many 
words what can be expressed with a few is meaningless. Thus, it is important to 
summarize an article using a small number of  meaningful words.  Bradford law
for the  scattering of articles over different journals is connected to the  success-breeds-success  (SBS) principle \cite{price76}: success in the past 
increases chances for some success in the future. For   example, a journal that
has been frequently consulted for some purpose is more likely to be read again, rather
than one of previously infrequent use. 
\par
In order to obtain the law of Zipf-Mandelbrot, we start from the following
version of Lotka  law : $n_x = C/(1+ x)^{1+\alpha}$, 
where $x$ is the scientist's productivity, $\alpha$ is a characteristic exponent, 
$C$ is a constant which in most cases is equal to the number of authors with the minimal 
productivity $x = 1$, i.e., to $n_1$. On the basis of this formula,   the number 
of scientists $r$   who are characterized by productivity
$x_r< x < k_{max}$ ($k_{max}$ is the maximal productivity of a scientist) reads
\begin{equation}\label{yule5}
r=\sum_{x=x_r}^{k_{max}} n_r \approx C \int_{x_r}^{k_{max}}\frac{dx}{x^{1+ \alpha}} =
\frac{C}{\alpha} \left(\frac{1}{x_r^{\alpha}}- \frac{1}{k_{max}^{\alpha}} \right).
\end{equation}
Depending on the value of $x_r$, $r$ can have values $1,2,3,\dots$ and in such a way the
scientists can be ranked.
If all scientists  of a scientific community working on the same topic are ranked in the 
order of the decrease of their productivity, the place of a
scientist who has written $x_r$ papers will be determined by his/her rank $r$. 
When the productivity of a scientist $x_r$ is found from Eq.(\ref{yule5}) as a function of 
rank $r$,   the relationship
\begin{equation}\label{zm}
x_r = \left(\frac{A}{r + B}\right)^{\gamma};\hskip.25cm 
A = (C/\alpha)^{1/\alpha}; \hskip.25cm B = C/(\alpha 
k_{\max}^{\alpha} ); \hskip.25cm  \gamma = 1/\alpha.
\end{equation}
This is the rank law of Zipf-Mandelbrot, which generalizes Zipf law:
$f(r) = c r^{-\beta}; r=1,2,3,\dots$, where $c$ and $\beta$ are parameters.
   Zipf law was discovered by counting words in books. If words in a book
are ranked in decreasing order according to their number of occurrences, then  Zipf  law
states that the number of occurrences of a word is inversely proportional to its rank $r$.
\par
Assuming that in Lotka  law the exponent takes the value $\alpha = 1$  and that in 
most cases $C = n_1$, one has  $x_r = n_1/(r + a)$, where 
$a = n_1/k_{max}$, $r \ge 0$. Integration of the last relationship yields the total productivity 
$R(n)$ of all scientists, beginning
with the one with the greatest productivity $k_{max}$ and ending with the scientist
whose productivity corresponds to the rank $n$ (the scientists are ranked in the
order of diminishing productivity; the rank is assumed to be a continuous-like variable):
\begin{equation}\label{bradford}
R(n) = n_1 \ln \left( \frac{n}{a} +1 \right).
\end{equation}
This is   Bradford  law.   According to this law,  for a given topic, a large
number of  relevant articles will be concentrated in a small number of journals.
The remaining articles will be dispersed over a large number of journals. Thus,
if scientific journals are arranged in order of
decreasing published articles on a given subject, they may be split to a
core of journals more particularly devoted to the subject and a shell consisting of sub-shells  of journals
containing the same numbers of articles as the core. Then  the number
of journals from the core zone and succeeding sub-shells will follow the relationship 
$1 : n : n^2 : \dots$.
\begin{center}
{\fbox{\fbox{\parbox{13.0cm}{ \vskip0.5cm \sf 
FOR POLICY-MAKERS\\
 Take away box Nr.15: \\The Zipf-Pareto  law, in the case of the distribution
 of scientists with
respect to  their productivity, indicates  that one can always single out a small number of  productive
scientists who wrote the greatest number of papers on a given subject, and
a large number of scientists with low productivity. The same applies also to  
scientific contacts, citation networks, etc.
This specific feature (so-called hierarchical stratification) of  the Zipf-Pareto law 
reflects a basic mechanism in the formation  of stable complex systems. 
This  can/must be taken into account in the process of
planning and the organization of science.  
 \vskip0.5cm }}}}\\
\end{center}
\section{Concluding remarks}
Knowledge has a complex nature. It can be created. It can lead to 
innovations and new technologies, and on this base, knowledge supports the advance 
and economic growth of  societies.  Knowledge can be collected.  Knowledge can be spread. 
Diffusion of ideas is closely connected to the
collection and spreading of knowledge. Some stages of the diffusion of ideas can be
described by epidemic models of scientific and technological systems. Most of
the  models described here are deterministic, but if
the internal and external fluctuations are strong, then different kinds of  models 
can be applied taking into account stochastic features.  

Much information about properties and stability of the knowledge systems
can be obtained by the statistical approach on the basis of distributions connected to
the Lotka-Volterra models of diffusion of knowledge. Interestingly, new terms occur in the usual  evolution equations because of the variability and flexibility in the opinions of actors, due to media contacts or interpersonal contacts, when exchanging ideas.

The inclusion of spatial variables
in the models leads to new research topics, such as questions on the spreading of systems of ideas
and competition among ideas in different areas/countries. 

In conclusion, the
epidemiological perspective renders a piece of mosaic to a better understanding of
the dynamics of diffusion of ideas in science, technology, and society, which  should be
  one of the main future tasks of the science of science \cite{wagner}.

\begin{flushleft}
{\large \bf Acknowledgment}
\end{flushleft}
Thanks to the editors of the book for inspiring us into writing this work. The authors gratefully acknowledge stimulating discussions with many wonderful colleagues at  several meetings of the ESF Action
COST MP-0801 'Physics of Competition and Conflict'. 


\begin{thebibliography}{99}
\bibitem{gordon} For example, at Gordon Research Conferences, it is forbidden to take written notes and to quote participant interventions later.
\end{thebibliography}
\end{document}